\documentclass[aps,prl,reprint, amsmath, amssymb, aps, superscriptaddress,longbibliography]{revtex4-2}

\usepackage{hyperref}
\hypersetup{colorlinks=true,linkcolor=blue,citecolor = blue, urlcolor=blue}

\usepackage{graphicx}
\usepackage{dcolumn}
\usepackage{bm}
\usepackage[utf8]{inputenc}
\usepackage{amssymb}
\usepackage{amsthm}
\usepackage{mathrsfs}
\usepackage{dsfont}
\usepackage{multirow}
\usepackage{enumitem}
\usepackage{stmaryrd} 
\usepackage{float}
\usepackage{physics}
\usepackage{cancel}
\usepackage{array}
\usepackage{longtable}
\usepackage{soul}
\usepackage{breakurl}

\usepackage{orcidlink}
\usepackage{tikzit}

\tikzstyle{red dot}=[fill=red, draw=none, shape=circle]
\tikzstyle{green dot}=[fill={rgb,255: red,70; green,176; blue,20}, draw=none, shape=circle]
\tikzstyle{Resize}=[font={\scriptsize}]
\tikzstyle{Big}=[font={\huge}]
\tikzstyle{Photon}=[fill={rgb,255: red,255; green,230; blue,103}, draw=none, shape=circle, minimum size=4pt]
\tikzstyle{Big node}=[fill={rgb,255: red,191; green,191; blue,191}, draw=black, shape=circle]
\tikzstyle{small node}=[fill={rgb,255: red,191; green,191; blue,191}, draw=black, shape=circle, inner sep=0, minimum size=300pt]
\tikzstyle{Blue dot}=[fill=blue, draw=black, shape=circle]
\tikzstyle{Black dot}=[fill=black, draw=none, shape=circle, inner sep=0pt, minimum size=4pt]

\tikzstyle{Fill red}=[-, fill={rgb,255: red,255; green,184; blue,184}]
\tikzstyle{dashes}=[-, dashed, dash pattern=on 1mm off 1mm]
\tikzstyle{Fill grey}=[-, fill={rgb,255: red,166; green,166; blue,166}]
\tikzstyle{Thick}=[-, thick]
\tikzstyle{Arrow}=[->]
\tikzstyle{red line}=[-, fill=none, draw=red]
\tikzstyle{Blue line}=[-, fill=none, draw=blue]
\tikzstyle{Gray line}=[-, fill=none, draw=gray]
\tikzstyle{Thick arrow}=[thick, ->]
\tikzstyle{Fill pink}=[-, fill={rgb,255: red,255; green,140; blue,140}]
\tikzstyle{dashed arrow}=[->, dashed, dash pattern=on 1mm off 0.5mm]
\tikzstyle{arrow}=[->, very thick, draw=red]
\tikzstyle{dashes red}=[-, dashed, dash pattern=on 1mm off 1mm, fill=none, draw=red]
\tikzstyle{dashes blue}=[-, dashed, dash pattern=on 1mm off 1mm, fill=none, draw=blue]
\tikzstyle{Thick dashed arrow}=[->, thick, dashed, dash pattern=on 1mm off 0.5mm]
\tikzstyle{Fill green}=[-, fill={rgb,255: red,70; green,176; blue,20}, draw=none]
\tikzstyle{Fill real red}=[-, fill=red, draw=none]
\tikzstyle{fill grey borderless}=[-, draw=none, fill={rgb,255: red,166; green,166; blue,166}]

\newcommand{\C}{\mathbb{C}}

\newcommand{\1}{\mathds{1}}

\newcommand{\mycomment}[1]{}
\newcommand{\vac}{\ket{\text{vac}}}

\newcommand{\changes}[1]{{#1}}

\newcounter{Show}
\begin{document}

\preprint{APS/123-QED}

\ifnum  2=\theShow
    \title{Supplemental Material for: The Role of Symmetry in Generalized Hong-Ou-Mandel Interference and Quantum Metrology}
\else
    \title{The Role of Symmetry in Generalized Hong-Ou-Mandel Interference and Quantum Metrology}
\fi

\author{\'Eloi Descamps\orcidlink{0000-0002-6911-452X}}
\email[Contact author: ]{eloi.descamps@u-paris.fr}
\affiliation{Universitée Paris Cité, CNRS, Laboratoire Matériaux et Phénomènes Quantiques, 75013 Paris, France}
\author{Arne Keller\orcidlink{0000-0002-6934-7198}}
\affiliation{Universitée Paris Cité, CNRS, Laboratoire Matériaux et Phénomènes Quantiques, 75013 Paris, France}
\affiliation{Département de Physique, Université Paris-Saclay, 91405 Orsay Cedex, France}
\author{Pérola Milman\orcidlink{0000-0002-7579-7742}}
\affiliation{Universitée Paris Cité, CNRS, Laboratoire Matériaux et Phénomènes Quantiques, 75013 Paris, France}

\date{\today}

\ifnum 2=\theShow\else
\begin{abstract}
    The Hong-Ou-Mandel interferometer is a foundational tool in quantum optics, with both fundamental and practical significance. Earlier works identified that input-state symmetry under exchange of the two spatial modes is fundamental in the understanding of the Hong-Ou-Mandel effect. We now show that this notion of symmetry is central to generalizing this effect. In particular, this point of view enables the construction of extensions beyond the standard two single-photon case to arbitrary input states, as well as to configurations with more than two spatial modes via a natural generalization of the beam splitter to a discrete Fourier transform interferometer. Beyond its conceptual significance, this framework offers direct insights into quantum metrology, showing how symmetry properties of input states allow the computation of explicit precision bounds. By focusing on symmetry, we provide a perspective that simplifies and unifies a range of known results, while paving the way for new developments in quantum interference and sensing.
\end{abstract}
\fi

\maketitle


\ifnum 2=\theShow\else

The Hong-Ou-Mandel (HOM) effect \cite{hong_measurement_1987} is one of the most fundamental manifestations of quantum interference in quantum optics. Initially identified as a dip in coincidence counts when two photons in indistinguishable modes impinge on a beam splitter (BS), the effect has become emblematic of the nonclassical behavior of bosons. Beyond its foundational significance \cite{douce_direct_2013}, HOM interference plays a central role in quantum technologies, with applications ranging from photonic quantum computing \cite{aaronson_computational_2013,knill_scheme_2001,fabre_hongoumandel_2022} to precision metrology \cite{chen_hong-ou-mandel_2019,fabre_parameter_2021,jordan_quantum_2022,lyons_attosecond-resolution_2018}. Despite its apparent simplicity, the HOM effect exhibits rich theoretical and experimental subtleties and has inspired numerous analyses and generalizations. At its core, the HOM effect concerns the mixing of two modes that are otherwise identical in all other degrees of freedom, which can nevertheless serve as path markers carrying which-way information; in the case considered here, these modes are spatial and are mixed via a BS.

Previous works have highlighted the roles of mode distinguishability \cite{tsujino_distinguishing_2004,ou_distinguishing_2005,dittel_wave-particle_2021,ma_unraveling_2025}, purity \cite{jones_distinguishability_2023}, and modal structure \cite{walborn_multimode_2003} in determining the degree of photon bunching. Extensions of the HOM effect beyond the case of independent single-photon inputs have also been extensively investigated. It has been shown, both theoretically and experimentally, that mode entanglement can give rise to anti-bunching \cite{walborn_multimode_2003,douce_direct_2013,eckstein_broadband_2008} and other nontrivial interference phenomena \cite{francesconi_anyonic_2021,chen_hong-ou-mandel_2019}. Generalizations to multi-photon inputs have revealed that the parity of the photon number in the input state plays a crucial role in the appearance of interference \cite{alsing_hong-ou-mandel_2024,alsing_extending_2022,chiruvelli_parity_2011}. Finally, with its single BS directly followed by detection, the HOM effect can be considered as the simplest linear interferometric effect, and adding spatial modes \cite{guanzon_multimode_2021}, beyond the two inputs of the BS, leads to setups akin to those considered for boson sampling tasks \cite{pioge_anomalous_2024, menssen_distinguishability_2017} for which permutation properties of the input state plays a fundamental role \cite{englbrecht_indistinguishability_2024}. 

Already in the original work by Hong, Ou, and Mandel \cite{hong_measurement_1987}, the potential of the HOM effect for precision parameter estimation was recognized. Subsequent research has further advanced this metrological capability \cite{lyons_attosecond-resolution_2018,chen_hong-ou-mandel_2019} and broadened its scope \cite{descamps_time-frequency_2023,meskine_approaching_2024}. Naturally, the various extensions mentioned above have been explored for their implications in quantum metrology, aiming to harness particle statistics \cite{sun_projection_2006,anisimov_quantum_2010,hofmann_all_2009,gao_super-resolution_2010, birrittella_multiphoton_2012, ben-aryeh_phase_2012, seshadreesan_phase_2013}, modal distributions \cite{chen_hong-ou-mandel_2019,descamps_quantum_2023,descamps_time-frequency_2023}, and spatial entanglement \cite{motes_linear_2015,motes_linear_2017} to enhance the precision of parameter estimation protocols. In a previous work \cite{descamps_time-frequency_2023}, we presented a general framework for understanding the HOM effect and its metrological applications, based on the symmetry under exchange of the two spatial modes at the BS. Building on this insight, we derived an explicit expression for the precision of the HOM interferometer across a broad class of configurations. 

The contribution of this Letter is two-fold: first, we demonstrate how the symmetry-based approach offers a compact and intuitive understanding of generalized HOM effects; second, we show how this framework enables the computation of metrological precision in a wide range of scenarios, allowing us to identify when generalized HOM interferometers can achieve the quantum precision limit. For pedagogical clarity, and given the significance of the two spatial modes case, we present our results with increasing generality. We begin by reviewing previously established results, then extend them to the important case of two spatial modes, and finally discuss the fully general scenario by considering an $n$-mode HOM-like interferometer. Throughout this letter, we consider only pure states. However, as we show in the supplemental material (SM) \cite{supmat}, all the results presented can be extended to mixed states.

The operator $\hat a_i^\dagger(\lambda)$ creates a photon in spatial mode $i \in \{0, \dots, n-1\}$ ($i = 0, 1$ corresponding to the two input ports of the HOM interferometer; see Fig.~\ref{fig:HOM}) and internal degree of freedom $\lambda$, which may represent any discrete or continuous property of the photon, such as frequency, polarization, or temporal profile. In this work, we first consider single-photon states which are superpositions of states of the form $\hat a_0^\dagger(\lambda_0) \cdots \hat a_{n-1}^\dagger(\lambda_{n-1}) \vac$, for arbitrary values of $\lambda_0, \dots, \lambda_{n-1}$, where $\vac$ denotes the vacuum state. For $n = 2$, this encompasses the standard input states of the HOM interferometer. More generally, we also consider states in which the photon number in each spatial mode is arbitrary and unconstrained (see SM \cite{supmat} for details).

\begin{figure*}
    \centering
    \begin{tabular}{ccc}
        \scalebox{0.8}{\begin{tikzpicture}
	\begin{pgfonlayer}{nodelayer}
		\node [style=none] (0) at (-1.5, 0) {};
		\node [style=none] (1) at (0, 1.5) {};
		\node [style=none] (2) at (1.5, 0) {};
		\node [style=none] (3) at (0, -1.5) {};
		\node [style=none] (4) at (-1, 1) {};
		\node [style=none] (5) at (1, 1) {};
		\node [style=none] (6) at (1, -1) {};
		\node [style=none] (7) at (3, 3) {};
		\node [style=none] (8) at (3, -3) {};
		\node [style=none] (9) at (3, 4) {};
		\node [style=none] (10) at (4, 3) {};
		\node [style=none] (11) at (4, -3) {};
		\node [style=none] (12) at (3, -4) {};
		\node [style=none] (13) at (4, 4) {};
		\node [style=none] (14) at (5.75, 4) {};
		\node [style=none] (15) at (5.75, 2.25) {};
		\node [style=none] (16) at (6, 1.25) {};
		\node [style=none] (17) at (4, -4) {};
		\node [style=none] (18) at (5.75, -3.75) {};
		\node [style=none] (19) at (7, -2.5) {};
		\node [style=none] (20) at (6, -1.25) {};
		\node [style=none] (21) at (0, -2.5) {BS};
		\node [style=none] (22) at (2.5, 1.5) {0};
		\node [style=none] (23) at (2.5, -1.5) {1};
		\node [style=none] (24) at (-3, 2) {};
		\node [style=Photon] (25) at (-3.75, 1.75) {~~~};
		\node [style=Photon] (26) at (-3.75, -1.75) {~~~};
		\node [style=none] (27) at (-2, 1.5) {0};
		\node [style=none] (28) at (-2, -1.5) {1};
		\node [style=none] (29) at (10, 0) {$P_c$};
		\node [style=none] (30) at (4, 1.25) {};
		\node [style=none] (31) at (8, 1.25) {};
		\node [style=none] (32) at (4, -1.25) {};
		\node [style=none] (33) at (8, -1.25) {};
		\node [style=none] (34) at (6, 0.5) {Concidence};
		\node [style=none] (35) at (6, -0.5) {detection};
		\node [style=none] (36) at (8.25, 0) {};
		\node [style=none] (37) at (9.5, 0) {};
		\node [style=none] (41) at (-1, -1) {};
		\node [style=none] (42) at (-3, -2) {};
		\node [style=none] (55) at (-4, 0.5) {};
		\node [style=none] (56) at (-3.5, 0.5) {};
		\node [style=none] (57) at (-3.75, 0.75) {};
		\node [style=none] (58) at (-3.75, 0) {};
		\node [style=none] (59) at (-4, -0.5) {};
		\node [style=none] (60) at (-3.5, -0.5) {};
		\node [style=none] (61) at (-3.75, -0.75) {};
		\node [style=none] (62) at (-4.75, 0) {$\ket{\psi}$};
		\node [style=none] (63) at (0, 2.25) {$\hat U$};
	\end{pgfonlayer}
	\begin{pgfonlayer}{edgelayer}
		\draw [style=Fill grey] (2.center)
			 to (3.center)
			 to (0.center)
			 to (1.center)
			 to cycle;
		\draw (0.center) to (2.center);
		\draw [style=Fill red] (10.center)
			 to (9.center)
			 to [bend left=90, looseness=1.75] cycle;
		\draw [style=Fill red] (12.center)
			 to (11.center)
			 to [bend left=90, looseness=1.75] cycle;
		\draw [in=165, out=45, looseness=1.75] (13.center) to (14.center);
		\draw [in=60, out=-15, looseness=1.25] (14.center) to (15.center);
		\draw [in=90, out=-105] (15.center) to (16.center);
		\draw [in=165, out=-45] (17.center) to (18.center);
		\draw [in=-60, out=-15, looseness=1.50] (18.center) to (19.center);
		\draw [in=-90, out=120, looseness=1.25] (19.center) to (20.center);
		\draw (30.center) to (31.center);
		\draw (31.center) to (33.center);
		\draw (33.center) to (32.center);
		\draw (32.center) to (30.center);
		\draw [style=Thick arrow] (36.center) to (37.center);
		\draw [style=dashed arrow] (5.center) to (7.center);
		\draw [style=dashed arrow] (6.center) to (8.center);
		\draw [style=dashed arrow, in=120, out=15, looseness=1.25] (24.center) to (4.center);
		\draw [style=dashed arrow, in=-120, out=-15, looseness=1.25] (42.center) to (41.center);
		\draw [bend left=45] (55.center) to (57.center);
		\draw [bend left=45, looseness=1.25] (57.center) to (56.center);
		\draw [in=90, out=-120, looseness=0.75] (58.center) to (59.center);
		\draw [bend right=45] (59.center) to (61.center);
		\draw [bend right=45, looseness=1.25] (61.center) to (60.center);
		\draw [in=-60, out=90, looseness=0.75] (60.center) to (58.center);
		\draw [in=-90, out=120, looseness=0.75] (58.center) to (55.center);
		\draw [in=60, out=-90, looseness=0.75] (56.center) to (58.center);
	\end{pgfonlayer}
\end{tikzpicture}} &~~~~ &\scalebox{0.8}{\begin{tikzpicture}
	\begin{pgfonlayer}{nodelayer}
		\node [style=none] (0) at (-1.5, 0) {};
		\node [style=none] (1) at (0, 1.5) {};
		\node [style=none] (2) at (1.5, 0) {};
		\node [style=none] (3) at (0, -1.5) {};
		\node [style=none] (4) at (-1, 1) {};
		\node [style=none] (5) at (1, 1) {};
		\node [style=none] (6) at (1, -1) {};
		\node [style=none] (7) at (3, 3) {};
		\node [style=none] (8) at (3, -3) {};
		\node [style=none] (9) at (3, 4) {};
		\node [style=none] (10) at (4, 3) {};
		\node [style=none] (11) at (4, -3) {};
		\node [style=none] (12) at (3, -4) {};
		\node [style=none] (13) at (4, 4) {};
		\node [style=none] (14) at (5.75, 4) {};
		\node [style=none] (15) at (5.75, 2.25) {};
		\node [style=none] (16) at (6, 1.25) {};
		\node [style=none] (17) at (4, -4) {};
		\node [style=none] (18) at (5.75, -3.75) {};
		\node [style=none] (19) at (7, -2.5) {};
		\node [style=none] (20) at (6, -1.25) {};
		\node [style=none] (21) at (0, -2.5) {BS};
		\node [style=none] (22) at (2.5, 1.5) {0};
		\node [style=none] (23) at (2.5, -1.5) {1};
		\node [style=none] (24) at (-3, 3) {};
		\node [style=Photon] (25) at (-7.5, 1.5) {~~~};
		\node [style=Photon] (26) at (-7.5, -1.5) {~~~};
		\node [style=none] (27) at (-2, 1.5) {0};
		\node [style=none] (28) at (-2, -1.5) {1};
		\node [style=none] (29) at (10, 0) {$P_c$};
		\node [style=none] (30) at (4, 1.25) {};
		\node [style=none] (31) at (8, 1.25) {};
		\node [style=none] (32) at (4, -1.25) {};
		\node [style=none] (33) at (8, -1.25) {};
		\node [style=none] (34) at (6, 0.5) {Concidence};
		\node [style=none] (35) at (6, -0.5) {detection};
		\node [style=none] (36) at (8.25, 0) {};
		\node [style=none] (37) at (9.5, 0) {};
		\node [style=none] (38) at (-4, -1) {$e^{i\kappa\hat H}$};
		\node [style=none] (39) at (-7, 2) {};
		\node [style=none] (40) at (-5, 3) {};
		\node [style=none] (41) at (-1, -1) {};
		\node [style=none] (42) at (-3, -3) {};
		\node [style=none] (43) at (-7, -2) {};
		\node [style=none] (44) at (-5, -3) {};
		\node [style=none] (45) at (-4.75, 3.5) {};
		\node [style=none] (46) at (-3.25, 3.5) {};
		\node [style=none] (47) at (-3.25, -3.5) {};
		\node [style=none] (48) at (-4.75, -3.5) {};
		\node [style=none] (49) at (-5, 3.25) {};
		\node [style=none] (50) at (-3, 3.25) {};
		\node [style=none] (51) at (-3, -3.25) {};
		\node [style=none] (52) at (-5, -3.25) {};
		\node [style=none] (53) at (-4, 1) {$\hat V(\kappa)$};
		\node [style=none] (54) at (-4, 0) {$=$};
		\node [style=none] (55) at (-7.75, 0.5) {};
		\node [style=none] (56) at (-7.25, 0.5) {};
		\node [style=none] (57) at (-7.5, 0.75) {};
		\node [style=none] (58) at (-7.5, 0) {};
		\node [style=none] (59) at (-7.75, -0.5) {};
		\node [style=none] (60) at (-7.25, -0.5) {};
		\node [style=none] (61) at (-7.5, -0.75) {};
		\node [style=none] (62) at (-8.5, 0) {$\ket{\psi}$};
		\node [style=none] (63) at (0, 2.25) {$\hat U$};
	\end{pgfonlayer}
	\begin{pgfonlayer}{edgelayer}
		\draw [style=Fill grey] (2.center)
			 to (3.center)
			 to (0.center)
			 to (1.center)
			 to cycle;
		\draw (0.center) to (2.center);
		\draw [style=Fill red] (10.center)
			 to (9.center)
			 to [bend left=90, looseness=1.75] cycle;
		\draw [style=Fill red] (12.center)
			 to (11.center)
			 to [bend left=90, looseness=1.75] cycle;
		\draw [in=165, out=45, looseness=1.75] (13.center) to (14.center);
		\draw [in=60, out=-15, looseness=1.25] (14.center) to (15.center);
		\draw [in=90, out=-105] (15.center) to (16.center);
		\draw [in=165, out=-45] (17.center) to (18.center);
		\draw [in=-60, out=-15, looseness=1.50] (18.center) to (19.center);
		\draw [in=-90, out=120, looseness=1.25] (19.center) to (20.center);
		\draw (30.center) to (31.center);
		\draw (31.center) to (33.center);
		\draw (33.center) to (32.center);
		\draw (32.center) to (30.center);
		\draw [style=Thick arrow] (36.center) to (37.center);
		\draw [style=dashed arrow] (5.center) to (7.center);
		\draw [style=dashed arrow] (6.center) to (8.center);
		\draw [style=dashed arrow, in=120, out=0, looseness=0.75] (24.center) to (4.center);
		\draw [style=dashed arrow, in=-180, out=45, looseness=0.75] (39.center) to (40.center);
		\draw [style=dashed arrow, in=180, out=-45, looseness=0.75] (43.center) to (44.center);
		\draw [style=dashed arrow, in=-120, out=0, looseness=0.75] (42.center) to (41.center);
		\draw (47.center)
			 to (48.center)
			 to [bend right=315] (52.center)
			 to (49.center)
			 to [bend left=45] (45.center)
			 to (46.center)
			 to [bend left=45] (50.center)
			 to (51.center)
			 to [bend left=45] cycle;
		\draw [bend left=45] (55.center) to (57.center);
		\draw [bend left=45, looseness=1.25] (57.center) to (56.center);
		\draw [in=90, out=-120, looseness=0.75] (58.center) to (59.center);
		\draw [bend right=45] (59.center) to (61.center);
		\draw [bend right=45, looseness=1.25] (61.center) to (60.center);
		\draw [in=-60, out=90, looseness=0.75] (60.center) to (58.center);
		\draw [in=-90, out=120, looseness=0.75] (58.center) to (55.center);
		\draw [in=60, out=-90, looseness=0.75] (56.center) to (58.center);
	\end{pgfonlayer}
\end{tikzpicture}}\\
        (a) & &(b)
    \end{tabular}  
    \caption{\changes{Schematics of the HOM interferometer. (a) Two photons in an arbitrary two-mode state $\ket{\psi}$ enter a balanced BS $\hat U$. Photon-number-resolving detectors at the output ports measure the number of photons exiting each port. The ports of the BS are labeled as $j=0,1$, referring to the indices of the creation operator $\hat a_j^\dagger(\lambda)$. (b) Metrological scenario: before entering the BS, the two-mode state $\ket{\psi}$ undergoes a parameter-dependent evolution $\hat V(\kappa) = e^{i \hat H \kappa}$, where $\kappa$ is the parameter to be estimated.}}
    \label{fig:HOM}
\end{figure*}
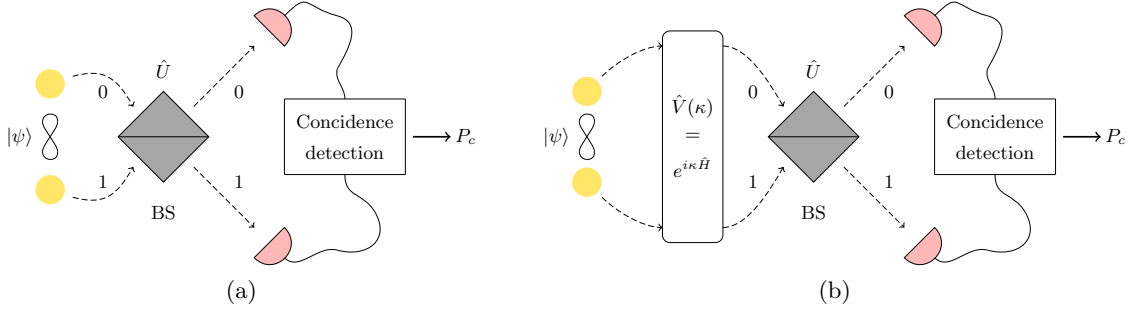

We consider the situation (see Fig.~\ref{fig:HOM} (a)) \cite{descamps_time-frequency_2023} where an arbitrary single-photon state $\ket{\psi}$ is input into a balanced BS $\hat U$ with the following transformation rule \changes{$\hat a_j^\dagger(\lambda)\mapsto\frac{1}{\sqrt{2}}\left(\hat a_0^\dagger(\lambda)+(-1)^j\hat a_1^\dagger(\lambda)\right)$ ($j=0,1$)}. Single-photon detectors at the output of the BS measure whether the two photons exit through the same port (bunching) or different ports (coincidence). We obtained the following compact expression for the coincidence probability:
\begin{equation}\label{eq: PC HOM single photon}
    P_c=\frac{1}{2}\left(1-\bra{\psi}\hat S\ket{\psi}\right),
\end{equation}
where the symmetry operator $\hat S$ swaps the two spatial modes. For any $\lambda$, $\hat S: \hat a_0^\dagger(\lambda)\leftrightarrow \hat a_1^\dagger(\lambda)$. This formula shows that, for pure states, the HOM effect is better understood as a measure of the symmetry of the modal distribution of the input state rather than the distinguishability of the incoming modes \cite{descamps_time-frequency_2023}.

In the considered metrological scenario, two photons arrive at the BS, and undergo a parameter-dependent evolution $\hat{V}(\kappa) = e^{i \hat{H} \kappa}$, where $\kappa$ is the parameter to be estimated \changes{and $\hat H$ is any Hermitian operator (see Fig.~\ref{fig:HOM} (b)).} The estimation precision $\delta\kappa$ is bounded by the Cramér-Rao and quantum Cramér-Rao bounds as
\begin{equation}\label{eq:CRB_QCRB}
    \delta\kappa \geq \frac{1}{\sqrt{N \mathcal{F}}} \geq \frac{1}{\sqrt{N \mathcal{Q}}},
\end{equation}
where $N$ is the number of independent measurements, $\mathcal{F}$ is the Fisher information (FI), and $\mathcal{Q}$ is the quantum Fisher information (QFI), which sets the ultimate precision limit attainable using any quantum measurement. For an experiment with possible outcomes $x \in X$, occurring with probability $P(x, \kappa)$, we have $\mathcal{F} = \sum_{x \in X} \frac{1}{P(x, \kappa)} \left( \frac{\partial P(x, \kappa)}{\partial \kappa} \right)^2$. For a pure state and the unitary evolution $e^{i\hat H\kappa}$, the QFI becomes $\mathcal{Q} = 4 \, \Delta \hat{H}$, where $\Delta \hat{H}$ denotes the variance of the generator $\hat{H}$ evaluated on the probe state.

If the initial probe state satisfies the symmetry condition $\hat{S} \ket{\psi} = \pm \ket{\psi}$, for estimation of the parameter near $\kappa=0$ \cite{descamps_time-frequency_2023}
\begin{equation}\label{eq: precision HOM single photon}
    \mathcal{F} = \Delta(\hat{H} - \hat{S} \hat{H} \hat{S}),
\end{equation}
where $\hat{S} \hat{H} \hat{S}$ represent $\hat{H}$ with the two spatial modes exchanged. This expression provides both a conceptual advantage by linking the symmetry properties of $\hat{H}$ and $\ket{\psi}$ to the precision of the HOM interferometer, and a practical one, by yielding an explicit formula for the FI in a wide variety of scenarios, not only limited to phase estimation. \changes{Crucially, it also highlights the importance of symmetries in metrology \cite{frerot_symmetry_2024}}.

We now show that Eqs.~(\ref{eq: PC HOM single photon}) and (\ref{eq: precision HOM single photon}) admit a generalization in the case where a two-spatial-mode state with a general photon number distribution is considered, and photon-number-resolving detection is performed after the BS. The core insight of our contribution is the following: since both the BS $\hat U$ and the symmetry operator $\hat S$ are passive linear optical operations, they can be described by  $2\times 2$ matrices (See SM \cite{supmat}).
\begin{align}
    \hat S: \begin{pmatrix} 0 & 1 \\ 1 & 0 \end{pmatrix}, \qquad \hat U: \frac{1}{\sqrt{2}} \begin{pmatrix} 1 & 1 \\ 1 & -1 \end{pmatrix}.
\end{align}

As this matrix representation is compatible with the composition of mode transformations, a matrix calculation shows that $\hat U^\dagger\hat S \hat U$ is represented by $\begin{pmatrix}1 & 0 \\ 0 & -1\end{pmatrix}$, which is the matrix representation of the parity operator $\hat \Pi_1$ on the mode labelled 1, meaning that
\begin{equation}\label{eq: relation sym parity}
    \hat U \hat S \hat U^\dagger = \hat \Pi_1.
\end{equation}
For states with a fixed number of photons, $\hat \Pi_1$ returns $+1$ for an even number of photons in mode 1, and $-1$ otherwise. The connection between the parity and the symmetry operator has already been partially observed in \cite{gao_super-resolution_2010} leading to the direct measurement the Wigner function \cite{douce_direct_2013}. However, to our knowledge, no analysis including the internal degrees of freedom has been performed, nor has $\hat S$ been explicitly identified as a symmetry operator. Equation~(\ref{eq: relation sym parity}) can be used to generalize (\ref{eq: PC HOM single photon}): if state $\ket{\psi}$ \changes{(not necessarily a single-photon state)} impinges on a BS, the probability $\mathbb P[n_1\equiv 0\pmod{2}]$ to measure an even number of photons in the spatial mode labeled 1 is
\begin{align}\label{eq: Peven HOM}
    \mathbb P[n_1\equiv 0\pmod{2}] &= \frac{1}{2} \left(1 + \bra{\psi} \hat U^\dagger \hat \Pi_1 \hat U \ket{\psi} \right)\notag\\
    &= \frac{1}{2} \left(1 + \bra{\psi} \hat S \ket{\psi} \right).
\end{align}
This equation reveals that parity is an important observable at the output of a BS when photon-number resolution is available \cite{birrittella_parity_2021, alsing_extending_2022, alsing_hong-ou-mandel_2024, chiruvelli_parity_2011}. \changes{Interestingly, by applying this formula to a product state, we recover the possibility to perform a continuous variable swap test with a beam splitter and parity measurement \cite{volkoff_ancilla-free_2022}.} Extending our previous result, the expression for $\mathbb P[n_1\equiv 0\pmod{2}]$ leads to a metrological protocol for which we can explicitly determine the precision. For an initial multi-photon state $\ket{\psi}$ and the evolution $e^{i\kappa \hat H}$, measuring the parity of the photon number in mode 1 after a BS allows the estimation of $\kappa$ with a precision limited by the Fisher information:
\begin{equation}\label{eq: precision extended HOM}
    \mathcal F = \Delta(\hat H - \hat S \hat H \hat S),
\end{equation}
if $\hat S \ket{\psi} = \pm \ket{\psi}$. As in Ref.~\cite{descamps_time-frequency_2023}, the formula is accurate only for parameter estimation around zero. However, if enough prior information is known on $\kappa$, an initial parameter shift can be performed such that the estimation is effectively done around the origin \cite{hervas_beyond_2025}. As this formula is identical to that of (\ref{eq: precision HOM single photon}), the same considerations about the symmetry of $\hat H$ apply: for anti-symmetric $\hat H$, the setup provides an optimal protocol. Furthermore, existing results \cite{anisimov_quantum_2010, ben-aryeh_phase_2012, birrittella_multiphoton_2012, chiruvelli_parity_2011}, can be reanalyzed from our point of view. Indeed, in most cases, the symmetry of the probe state can be assessed with minimal computation, automatically guaranteeing the optimality of the protocol. It is important to note that a related notion of symmetry, termed path symmetry, has been introduced in \cite{hofmann_all_2009, seshadreesan_phase_2013}. Similar to our work, those authors show that all path-symmetric states allow for optimal estimation protocols. However, path symmetry is an abstract concept, corresponding to a nonphysical exchange of the modes, distinct from the symmetry under exchange of spatial modes considered here. Another key difference is the implementation of the BS, and a more complete discussion on the impact of the choice of balanced BS can be found in \cite{supmat}. Finally, as the class of states for which we can compute the metrological precision has been extended, our formula offers promising avenues for enhanced parameter estimation tasks. Indeed, the variance $\Delta(\hat H - \hat S \hat H \hat S)$ can be increased using photon numbers beyond single-photon states. As our result holds for any photon number, a promising platform for experimentally verifying equation~(\ref{eq: Peven HOM}), using this effect to probe properties of multi-mode, multi-photon states and perform photon-number-enhanced metrology, is by spontaneous parametric down-conversion sources (SPDC) \cite{boucher_toolbox_2015, meskine_approaching_2024} operated out in a medium power pulsed-pump regime. In such a regime, a limited number of photon pairs are simultaneously generated, allowing photon-number resolution with existing photon-number-resolving technologies.

The ideas presented earlier can be extended to a more general setting by increasing the number of spatial modes. Fix an integer $n \geq 2$, and let $j\in\{0,\dots,n-1\}$ in $\hat a_j^\dagger(\lambda)$. To maintain our symmetry-based approach, we must generalize the operator $\hat S$. As discussed in the SM \cite{supmat}, there are multiple choices; however, up to permutations of the modes, the only meaningful one is defining $\hat P$ as the mode transformation
\begin{equation}
    \hat P \, \hat a_j^\dagger(\lambda) \, \hat P^\dagger = \hat a_{j-1}^\dagger(\lambda),
\end{equation}
where $j-1$ is taken modulo $n$. $\hat P$ performs a cyclic permutation of the spatial modes. The matrix associated with this linear optical transformation is the permutation matrix
\begin{equation}
    P = \scalebox{0.8}{$\begin{pmatrix} 
        0 & 1 & & \\
        & \ddots & \ddots & \\
        & & \ddots & 1 \\
        1 & & & 0 
    \end{pmatrix}$}.
\end{equation} 
Such matrices can be diagonalized (see \cite{supmat} for computational details) via the discrete Fourier transform (DFT) matrix $U$, with $U_{k,l} = \frac{1}{\sqrt{n}} \omega^{kl}$, where $\omega = e^{2\pi i/n}$. Consequently, denoting $\hat U$ the linear optical DFT interferometer~\cite{motes_linear_2015}, we have $\hat P = \hat U \hat D \hat U^\dagger$, where $\hat D$ implement $\hat a_i^\dagger(\lambda) \mapsto \omega^i \hat a_i^\dagger(\lambda)$. Analogous to how a BS splitter maps symmetry into parity, this decomposition implies that symmetry under cyclic permutation of modes is mapped, through the DFT interferometer, to a (complex) observable diagonal in the photon-number basis. Its properties are thus directly related to photon number statistics. We are therefore led to consider the following setup: an arbitrary $n$-mode spatial state $\ket{\psi}$ is input into a DFT interferometer $\hat U$, and photon-number-resolving detectors are placed at the output ports to measure the number $m_k$ of photons in mode $k$ (see Fig.~\ref{fig: fourier}). By combining the resulting terms (see SM \cite{supmat} for details), we can generalize Eq.~(\ref{eq: Peven HOM}) as
\begin{equation}\label{eq: proba fourier}
    \mathbb{P}\left[\sum_{k=0}^{n-1} k m_k \equiv 0 \pmod{n} \right] = \frac{1}{n} \sum_{l=0}^{n-1} \bra{\psi} \hat P^l \ket{\psi}.
\end{equation}
\begin{figure}
    \centering
    \scalebox{1}{\begin{tikzpicture}
	\begin{pgfonlayer}{nodelayer}
		\node [style=none] (0) at (1.25, 0) {$\hat U$};
		\node [style=none] (1) at (0.25, -3) {};
		\node [style=none] (2) at (2.25, -3) {};
		\node [style=none] (3) at (0.25, 3) {};
		\node [style=none] (4) at (2.25, 3) {};
		\node [style=none] (5) at (0.25, 2.25) {};
		\node [style=none] (6) at (0.25, 1.25) {};
		\node [style=none] (7) at (0.25, -2.25) {};
		\node [style=none] (8) at (0.25, -1.25) {};
		\node [style=none] (9) at (-1.25, 2.25) {};
		\node [style=none] (10) at (-1.25, 1.25) {};
		\node [style=none] (11) at (-1.25, -1.25) {};
		\node [style=none] (12) at (-1.25, -2.25) {};
		\node [style=none] (13) at (-0.5, 0.25) {$\vdots$};
		\node [style=none] (14) at (3.25, 2.25) {};
		\node [style=none] (15) at (3.25, 1.25) {};
		\node [style=none] (16) at (3.25, -2.25) {};
		\node [style=none] (17) at (3.25, -1.25) {};
		\node [style=none] (18) at (2.25, 2.25) {};
		\node [style=none] (19) at (2.25, 1.25) {};
		\node [style=none] (20) at (2.25, -1.25) {};
		\node [style=none] (21) at (2.25, -2.25) {};
		\node [style=none] (22) at (2.75, 0.25) {$\vdots$};
		\node [style=none] (23) at (4, 2.5) {};
		\node [style=none] (24) at (4, 2) {};
		\node [style=none] (25) at (4.5, 2.25) {};
		\node [style=none] (26) at (5, 2.25) {};
		\node [style=none] (27) at (5.75, 2.25) {$m_0$};
		\node [style=none] (28) at (4, 1.5) {};
		\node [style=none] (29) at (4, 1) {};
		\node [style=none] (30) at (4.5, 1.25) {};
		\node [style=none] (31) at (5, 1.25) {};
		\node [style=none] (32) at (5.75, 1.25) {$m_1$};
		\node [style=none] (37) at (6, -1.25) {$m_{n-2}$};
		\node [style=none] (42) at (6, -2.25) {$m_{n-1}$};
		\node [style=none] (43) at (-5.75, 0) {$\ket{\psi}$};
		\node [style=none] (44) at (-4.5, 0) {\scalebox{1}[8]{$\{$}};
		\node [style=none] (45) at (-2.25, 0) {$e^{i\kappa\hat H}$};
		\node [style=none] (46) at (-3, -3) {};
		\node [style=none] (47) at (-1.5, -3) {};
		\node [style=none] (48) at (-3, 3) {};
		\node [style=none] (49) at (-1.5, 3) {};
		\node [style=none] (50) at (-3.25, 2.25) {};
		\node [style=none] (51) at (-3.25, 1.25) {};
		\node [style=none] (52) at (-3.25, -2.25) {};
		\node [style=none] (53) at (-3.25, -1.25) {};
		\node [style=none] (54) at (-4.25, 2.25) {};
		\node [style=none] (55) at (-4.25, 1.25) {};
		\node [style=none] (56) at (-4.25, -1.25) {};
		\node [style=none] (57) at (-4.25, -2.25) {};
		\node [style=none] (58) at (-3.75, 0.25) {$\vdots$};
		\node [style=none] (59) at (4, -1) {};
		\node [style=none] (60) at (4, -1.5) {};
		\node [style=none] (61) at (4.5, -1.25) {};
		\node [style=none] (62) at (5, -1.25) {};
		\node [style=none] (63) at (4, -2) {};
		\node [style=none] (64) at (4, -2.5) {};
		\node [style=none] (65) at (4.5, -2.25) {};
		\node [style=none] (66) at (5, -2.25) {};
		\node [style=none] (67) at (-3.25, 2.75) {};
		\node [style=none] (68) at (-1.25, 2.75) {};
		\node [style=none] (69) at (-1.25, -2.75) {};
		\node [style=none] (70) at (-3.25, -2.75) {};
		\node [style=none] (71) at (-4, -0.5) {};
	\end{pgfonlayer}
	\begin{pgfonlayer}{edgelayer}
		\draw [style=Fill grey] (2.center)
			 to (1.center)
			 to (3.center)
			 to (4.center)
			 to cycle;
		\draw (9.center) to (5.center);
		\draw (6.center) to (10.center);
		\draw (11.center) to (8.center);
		\draw (7.center) to (12.center);
		\draw (18.center) to (14.center);
		\draw (15.center) to (19.center);
		\draw (20.center) to (17.center);
		\draw (16.center) to (21.center);
		\draw [style=Fill pink] (24.center)
			 to (23.center)
			 to [in=90, out=0, looseness=0.75] (25.center)
			 to [in=0, out=-90, looseness=0.75] cycle;
		\draw (25.center) to (26.center);
		\draw [style=Fill pink] (29.center)
			 to (28.center)
			 to [in=90, out=0, looseness=0.75] (30.center)
			 to [in=0, out=-90, looseness=0.75] cycle;
		\draw (30.center) to (31.center);
		\draw (68.center)
			 to [bend right=45, looseness=1.25] (49.center)
			 to (48.center)
			 to [bend right=45, looseness=1.25] (67.center)
			 to (70.center)
			 to [bend right=45] (46.center)
			 to (47.center)
			 to [bend right=45] (69.center)
			 to cycle;
		\draw (54.center) to (50.center);
		\draw (51.center) to (55.center);
		\draw (56.center) to (53.center);
		\draw (52.center) to (57.center);
		\draw [style=Fill pink] (60.center)
			 to (59.center)
			 to [in=90, out=0, looseness=0.75] (61.center)
			 to [in=0, out=-90, looseness=0.75] cycle;
		\draw (61.center) to (62.center);
		\draw [style=Fill pink] (64.center)
			 to (63.center)
			 to [in=90, out=0, looseness=0.75] (65.center)
			 to [in=0, out=-90, looseness=0.75] cycle;
		\draw (65.center) to (66.center);
	\end{pgfonlayer}
\end{tikzpicture}}
    \caption{Schematic of the DFT interferometer. \changes{An arbitrary $n$-mode state $\ket{\psi}$ enters a DFT interferometer $\hat U$. Photon-number-resolving detectors at the output ports measure the number of photons $m_0, \dots, m_{n-1}$ exiting each port. The quantity $\sum_{k=0}^{n-1} k m_k$ is then computed modulo $n$. The evolution layer $e^{i\kappa\hat H}$ is added in the metrological scenario.}}
    \label{fig: fourier}
\end{figure}
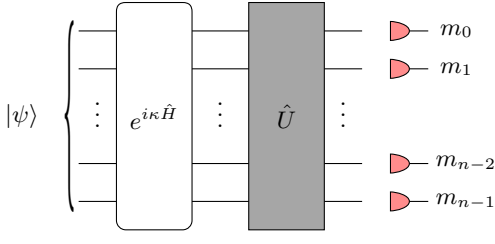
We can verify that Eq.~(\ref{eq: proba fourier}) reduces to Eq.~(\ref{eq: Peven HOM}) when $n=2$. As before, this formula implies that the symmetry properties of a state with respect to cyclic permutations can be probed by sending it through a DFT interferometer and selecting the outcomes for which $\sum_{k=0}^{n-1} k m_k \equiv 0 \pmod{n}$. Furthermore, as shown in \cite{supmat}, the operator $\hat \Pi = \frac{1}{n} \sum_{l=0}^{n-1} \hat P^l$ is the projector onto the eigenspace of $\hat P$ associated with eigenvalue $+1$. Therefore, the probability in Eq.~(\ref{eq: proba fourier}) evaluates the squared norm of the component of $\ket{\psi}$ in the $+1$ eigenspace of $\hat P$. Similarly, the probabilities corresponding to other values of the sum $\sum_{k=0}^{n-1} k m_k$ similarly probe the components of $\ket{\psi}$ in the other eigenspaces of $\hat P$ \cite{supmat}. \changes{This result is related to recent works studying the role of symmetry as observable for boson sampling problems \cite{shchesnovich_universality_2016,shchesnovich_partial_2015} and suppression laws in DFT interferometers\cite{tichy_zero-transmission_2010,dittel_totally_2018}.}

The interferometer under consideration can be used as a parameter estimation device. An initial probe state $\ket{\psi}$ undergo an evolution $e^{i\hat H\kappa}$, passes through the DFT interferometer, and the photon number distribution $m_0,\dots,m_{n-1}$ is recorded. We consider only two measurement outcomes: whether the quantity $\sum_{k=0}^{n-1} km_k$ is a multiple of $n$, or not (see Fig.~\ref{fig: fourier}). Interestingly, such an interferometer can be implemented in practice with a reasonable $O(n^2)$ amount of optical elements, as shown in \cite{reck_experimental_1994}. In \cite{supmat}, we show how to compute the FI for estimations around $\kappa=0$, and derive various expressions. Assuming that the initial state satisfies either $\hat \Pi\ket{\psi}=\ket{\psi}$ or $\hat \Pi\ket{\psi}=0$, we find that $\mathcal F = 4\Delta(i[\hat H,\hat \Pi])$, which characterizes, in the Heisenberg picture, how much the evolution generated by $\hat H$ changes the expectation value of $\hat \Pi$. However, it is generally difficult to assign a clear physical interpretation to $i[\hat H,\hat \Pi]$. Therefore, we provide two alternative expressions
\begin{equation}\label{eq: FI Fourier}
    \begin{split}
        \mathcal F= \frac{4}{n^2} \Delta\Big(n \hat{H} - \sum_{l=0}^{n-1} \hat{P}^l \hat{H} \hat{P}^{-l} \Big), \\
        \mathcal F=\frac{4}{n^2} \Delta\Big(\sum_{l=0}^{n-1} (-1)^l\hat{P}^l \hat{H} \hat{P}^{-l}\Big),
    \end{split}
\end{equation}
where the first equality holds if $\hat P\ket{\psi}=\ket{\psi}$ and the second if $\hat P\ket{\psi}=-\ket{\psi}$, which is possible only when $n$ is even. Since $\hat P^l \hat H \hat P^{-l}$ can be interpreted as $\hat H$ with the mode labels cyclically shifted $l$ times, both operators in Eq.~(\ref{eq: FI Fourier}) acquire a clear physical meaning. Comparing the FI to the QFI, $\mathcal Q = 4\Delta(\hat H)$, allows one to assess the optimality of the measurement. In particular, symmetry properties of $\hat H$ can suffice to determine optimality. For example, if the generator is anti-symmetric under permutation, i.e., $\hat P \hat H \hat P^\dagger = -\hat H$, then we directly have $\mathcal F = \mathcal Q$ in both cases.

A crucial assumption for our result to hold is that the initial state must satisfy a strong symmetry condition ($\hat P\ket{\psi}=\pm\ket{\psi}$). While such a condition can be met in simple situations, such as with identical independent single photons or entangled biphotons generated by SPDC sources \cite{boucher_toolbox_2015}, the experimental complexity of our protocol is justified only when more complex probe states are considered, specifically, those that yield a quantum advantage in the value of the QFI. Without delving into the details of generating highly entangled multimode states, a topic that remains an active area of research \cite{bensemhoun_multipartite_2025}, there exists a natural way to impose the required symmetry condition. The engineering of such states can be done via the insertion of an additional DFT interferometer before the evolution $e^{i\hat H\kappa}$. If the photon number distribution $\{m_k'\}$ of the incoming state is controlled such that $\sum_{k=0}^{n-1} km_k' \equiv 0 \pmod{n}$, then the symmetry of the state immediately before the evolution is guaranteed. This condition is, for example, satisfied when a single-photon state is used as input \footnote{In this case, $m_k'=1$, so $\sum_{k=0}^{n-1} k$ is congruent to $n/2$ or $0$ depending on the parity of $n$, which leads to a negative or positive symmetry, respectively.}. This results in an architecture reminiscent of a Mach-Zehnder interferometer.

To illustrate the versatility of our framework, we briefly discuss how it provides an explicit measurement setup capable of reaching the QFI studied in \cite{descamps_quantum_2023}. This study considers the metrological properties of $n$ single photons, each distributed in a different spatial mode. The role of frequency as an internal degree of freedom, and in particular, frequency entanglement, is investigated for time-delay estimation. The analysis uses the QFI as a figure of merit, which in this case is given by $\mathcal Q = 4\Delta\hat \Omega$, where $\hat \Omega = \sum_{l=0}^{n-1} \hat \omega_l$ is the collective time-delay generator, and $\hat \omega_l$ is the local time-evolution generator associated with mode $l$. The states considered in \cite{descamps_quantum_2023} can easily be made symmetric or antisymmetric, so, the interferometer we propose is well-suited to measure such evolutions. However, the operator $\hat \Omega$ is symmetric under the action of $\hat P$, and equation (\ref{eq: FI Fourier}) implies that $\mathcal F = 0$. To remedy this and instead achieve an optimal measurement, one needs to work with the operator $\hat \Omega' = \sum_{l=0}^{n-1} (-1)^l \hat \omega_l$, that is antisymmetric when $n$ is even, leading to $\mathcal F = \mathcal Q$. The analysis performed in the paper can be naturally adapted to the operator $\hat \Omega'$, thus yielding an explicit setup capable of measuring the QFI. \changes{Explicitly implementing the evolution generated by $\hat \Omega'$ is experimentally challenging for requiring to coordinnate positive and negative delays on different spatial modes. To circumvent this, one can use technics similar to those used in \cite{guanzon_multimode_2021, motes_linear_2015, motes_linear_2017} by
applying a delay to half of the modes, which up to global delay (which is irrelevant for interferometry) is equivalent to the previous scenario. This extends their range of applications by generalizing it to any degree of freedom. If $\hat P\ket{\psi}=\ket{\psi}$, the precision is given by}
\begin{equation}\label{eq: FI time delay}
    \changes{\mathcal F = 4 \Delta\left(\sum_{l=0}^{n/2-1} \hat \omega_l - \sum_{l=n/2}^{n-1} \hat \omega_l\right).}
\end{equation}
\changes{While Eq.~(\ref{eq: FI time delay}) does not necessarily equals the QFI, a proper choice of the input state will lead to a measurement precision exceeding the shot noise (experimental details are summarized in Fig.~\ref{fig: explicit exp}).}
\begin{figure}
    \centering
    \scalebox{0.89}{\begin{tikzpicture}
	\begin{pgfonlayer}{nodelayer}
		\node [style=none] (0) at (0.625, -1) {$\hat U$};
		\node [style=none] (1) at (-0.125, -3.5) {};
		\node [style=none] (2) at (1.375, -3.5) {};
		\node [style=none] (3) at (-0.125, 3.5) {};
		\node [style=none] (4) at (1.375, 3.5) {};
		\node [style=none] (5) at (-0.125, 2.75) {};
		\node [style=none] (6) at (-0.125, 0.5) {};
		\node [style=none] (7) at (-0.125, -2.75) {};
		\node [style=none] (8) at (-0.125, -0.75) {};
		\node [style=none] (9) at (-0.875, 2.75) {};
		\node [style=none] (10) at (-0.875, 0.5) {};
		\node [style=none] (12) at (-3.375, -2.75) {};
		\node [style=none] (14) at (2.125, 2.75) {};
		\node [style=none] (15) at (2.125, 0.5) {};
		\node [style=none] (16) at (2.125, -2.75) {};
		\node [style=none] (17) at (2.125, -0.75) {};
		\node [style=none] (18) at (1.375, 2.75) {};
		\node [style=none] (19) at (1.375, 0.5) {};
		\node [style=none] (20) at (1.375, -0.75) {};
		\node [style=none] (21) at (1.375, -2.75) {};
		\node [style=none] (22) at (1.75, -1.5) {$\vdots$};
		\node [style=none] (23) at (2.5, 3) {};
		\node [style=none] (24) at (2.5, 2.5) {};
		\node [style=none] (25) at (3, 2.75) {};
		\node [style=none] (26) at (3.5, 2.75) {};
		\node [style=none] (27) at (4.25, 2.75) {\scalebox{0.8}{$m_0$}};
		\node [style=none] (28) at (2.5, 0.75) {};
		\node [style=none] (29) at (2.5, 0.25) {};
		\node [style=none] (30) at (3, 0.5) {};
		\node [style=none] (31) at (3.5, 0.5) {};
		\node [style=none] (43) at (-7.375, -1.25) {$\ket{\psi}$};
		\node [style=none] (44) at (-6.125, 0) {\scalebox{1}[8]{$\{$}};
		\node [style=none] (46) at (-2.375, 0) {};
		\node [style=none] (47) at (-1.125, 0) {};
		\node [style=none] (48) at (-2.375, 3.5) {};
		\node [style=none] (49) at (-1.125, 3.5) {};
		\node [style=none] (50) at (-2.625, 2.75) {};
		\node [style=none] (51) at (-2.625, 0.5) {};
		\node [style=none] (54) at (-3.375, 2.75) {};
		\node [style=none] (55) at (-3.375, 0.5) {};
		\node [style=none] (56) at (-3.375, -0.75) {};
		\node [style=none] (58) at (-1.75, -1.5) {$\vdots$};
		\node [style=none] (59) at (2.5, -0.5) {};
		\node [style=none] (60) at (2.5, -1) {};
		\node [style=none] (61) at (3, -0.75) {};
		\node [style=none] (62) at (3.5, -0.75) {};
		\node [style=none] (63) at (2.5, -2.5) {};
		\node [style=none] (64) at (2.5, -3) {};
		\node [style=none] (65) at (3, -2.75) {};
		\node [style=none] (66) at (3.5, -2.75) {};
		\node [style=none] (67) at (-2.625, 3.25) {};
		\node [style=none] (68) at (-0.875, 3.25) {};
		\node [style=none] (69) at (-0.875, 0.25) {};
		\node [style=none] (70) at (-2.625, 0.25) {};
		\node [style=none] (72) at (-1.75, 2.75) {\scalebox{0.8}{Delay}};
		\node [style=none] (73) at (-1.75, 1.25) {$\kappa$};
		\node [style=none] (74) at (-0.5, 1.75) {$\vdots$};
		\node [style=none] (75) at (1.75, 1.75) {$\vdots$};
		\node [style=none] (76) at (-3, 1.75) {$\vdots$};
		\node [style=none] (78) at (-7.375, 0.25) {\scalebox{0.8}{State}};
		\node [style=none] (79) at (-7.375, 1) {\scalebox{0.8}{Photon}};
		\node [style=none] (80) at (-7.375, 1.75) {\scalebox{0.8}{Single}};
		\node [style=none] (83) at (7.875, 1.25) {\scalebox{0.8}{Estimation of}};
		\node [style=none] (84) at (7.875, -0.25) {\scalebox{0.7}{$\mathbb P\!\left[\sum\limits_{k=0}^{2n-1} k m_k\equiv 0 \!\pmod{2n}\!\right]$}};
		\node [style=none] (85) at (5, 2.25) {};
		\node [style=none] (86) at (10.75, 2.25) {};
		\node [style=none] (87) at (5, -1.25) {};
		\node [style=none] (88) at (10.75, -1.25) {};
		\node [style=none] (89) at (7.875, -2) {};
		\node [style=none] (90) at (7.875, -1.25) {};
		\node [style=none] (91) at (7.875, -2.5) {$\kappa$};
		\node [style=none] (94) at (4.25, 0.5) {\scalebox{0.8}{$m_{n-1}$}};
		\node [style=none] (95) at (4.25, -0.75) {\scalebox{0.8}{$m_n$}};
		\node [style=none] (96) at (4.25, -2.75) {\scalebox{0.8}{$m_{2n-1}$}};
		\node [style=none] (97) at (-4.125, -1) {$\hat U$};
		\node [style=none] (98) at (-4.875, -3.5) {};
		\node [style=none] (99) at (-3.375, -3.5) {};
		\node [style=none] (100) at (-4.875, 3.5) {};
		\node [style=none] (101) at (-3.375, 3.5) {};
		\node [style=none] (103) at (-4.875, 2.75) {};
		\node [style=none] (104) at (-4.875, 0.5) {};
		\node [style=none] (105) at (-5.75, 2.75) {};
		\node [style=none] (106) at (-5.75, 0.5) {};
		\node [style=none] (109) at (-5.375, 1.75) {$\vdots$};
		\node [style=none] (110) at (-5.375, -1.5) {$\vdots$};
		\node [style=none] (111) at (-4.875, -2.75) {};
		\node [style=none] (112) at (-4.875, -0.75) {};
		\node [style=none] (113) at (-5.75, -2.75) {};
		\node [style=none] (114) at (-5.75, -0.75) {};
		\node [style=none] (115) at (-4.125, 0.75) {\scalebox{0.8}{DFT}};
		\node [style=none] (116) at (-4.125, 4) {\scalebox{0.65}{Optionnal}};
		\node [style=none] (117) at (2.25, -1.25) {};
		\node [style=none] (118) at (0.625, 0.75) {\scalebox{0.8}{DFT}};
	\end{pgfonlayer}
	\begin{pgfonlayer}{edgelayer}
		\draw [style=Fill grey] (2.center)
			 to (1.center)
			 to (3.center)
			 to (4.center)
			 to cycle;
		\draw (9.center) to (5.center);
		\draw (6.center) to (10.center);
		\draw (7.center) to (12.center);
		\draw (18.center) to (14.center);
		\draw (15.center) to (19.center);
		\draw (20.center) to (17.center);
		\draw (16.center) to (21.center);
		\draw [style=Fill pink] (24.center)
			 to (23.center)
			 to [in=90, out=0, looseness=0.75] (25.center)
			 to [in=0, out=-90, looseness=0.75] cycle;
		\draw (25.center) to (26.center);
		\draw [style=Fill pink] (29.center)
			 to (28.center)
			 to [in=90, out=0, looseness=0.75] (30.center)
			 to [in=0, out=-90, looseness=0.75] cycle;
		\draw (30.center) to (31.center);
		\draw (68.center)
			 to [bend right=45, looseness=1.25] (49.center)
			 to (48.center)
			 to [bend right=45, looseness=1.25] (67.center)
			 to (70.center)
			 to [bend right=45] (46.center)
			 to (47.center)
			 to [bend right=45] (69.center)
			 to cycle;
		\draw (54.center) to (50.center);
		\draw (51.center) to (55.center);
		\draw [style=Fill pink] (60.center)
			 to (59.center)
			 to [in=90, out=0, looseness=0.75] (61.center)
			 to [in=0, out=-90, looseness=0.75] cycle;
		\draw (61.center) to (62.center);
		\draw [style=Fill pink] (64.center)
			 to (63.center)
			 to [in=90, out=0, looseness=0.75] (65.center)
			 to [in=0, out=-90, looseness=0.75] cycle;
		\draw (65.center) to (66.center);
		\draw (56.center) to (8.center);
		\draw (85.center) to (86.center);
		\draw (88.center) to (86.center);
		\draw (88.center) to (87.center);
		\draw (87.center) to (85.center);
		\draw [style=Thick arrow] (90.center) to (89.center);
		\draw [style=fill grey borderless] (99.center)
			 to (98.center)
			 to (100.center)
			 to (101.center)
			 to cycle;
		\draw [style=dashes] (100.center)
			 to (101.center)
			 to (99.center)
			 to (98.center)
			 to cycle;
		\draw (105.center) to (103.center);
		\draw (104.center) to (106.center);
		\draw (111.center) to (113.center);
		\draw (114.center) to (112.center);
	\end{pgfonlayer}
\end{tikzpicture}}
    \caption{\changes{Proposal for the measurent of a collective delay unsing single photon state. An initial single-photon state is prepared, for example using SPDC sources. Optionnaly, it can be sent through a DFT interferometer $\hat U$ to impose the required symmetry. The parameter-dependent evolution $\hat V(\kappa)$ is implemented by applying time delays to half of the modes. Finally, another DFT interferometer is used before photon-number-resolving detection to extract the relevant measurement statistics.}}
    \label{fig: explicit exp}
\end{figure}
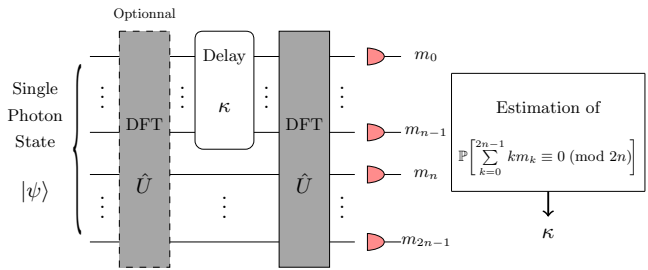
We now briefly discuss the role and effects of imperfections. First, it is experimentally impossible to produce states that exactly satisfy the symmetry conditions required for our computation of the FI to hold. This limitation is a generalization of a problem already present in HOM metrology, and in \cite{meskine_approaching_2024} we presented a general approach to address it: \changes{the symmetry imperfections reduce the visibility of the HOM interferometer and by proposing a model for the visibility dependence, we observed that the precision (maximal attainable FI for this specfic measurement strategy) decreases quadraticaly with the visibility.} This approach can be adapted to our present contribution.

\changes{We now discuss photon loss. Since the protocol relies on accurate photon-number statistics to evaluate $\sum_{k=0}^{n-1} k m_k$, photon losses alter the outcome probabilities. Assuming a fixed number of photons, post-selection on events where all photons are detected, can be used. While it becomes inefficient as losses scale exponentially with photon number, under realistic experimental efficiencies ($\eta \gtrsim 0.9$), protocols involving up to ten photons are within reach of current photonic platforms~\cite{ding_photon-number-resolving_2025}. In \cite{supmat}, we have developed further tools to help analyze the effect of losses on the precision and its relation to state preparation.}

\changes{Finally, while our analysis has been presented in the context of photonic systems, the underlying principles can be extended to any bosonic one. As superconducting microwave cavities in circuit QED \cite{lang_correlations_2013}, quasi particle such as phonons \cite{desmarchelier_phonon_2022}, or magnon excitations in solid state systems \cite{song_single-shot_2025} lead to the observation of HOM-like effects, exploring our framework and protocol in these alternative platforms, could open new perspectives for quantum sensing and more broadly for quantum systems manipulation and control.}
 
In conclusion, we have presented a general analysis that encapsulates much of the previous works within a unified picture inspired by a novel and fruitful point of view for studying the HOM interferometer using a symmetry-based approach. Secondly, it led to a generalization of the experimental setup to a more complex scenario, offering new and promising avenues for developing and testing metrological protocols with quantum optical systems. A natural next step is the design and implementation of a concrete experimental realization. We have presented a way to produce the states with the desired symmetry, and the required tailored photon states can, in principle, be produced with current technology. Such an experiment would not only demonstrate the metrological capabilities of the proposed interferometer but also showcase its broader versatility: beyond precision measurements, it could serve as a powerful diagnostic tool to probe and characterize the quantum state of light, in a manner reminiscent of how the HOM effect is routinely employed in quantum optics.

\section{Acknowledgments}\label{section: acknoledgments}
We acknowledge funding from the Plan France 2030
through the projects ANR-22-PETQ-0006 and ANR: NR-24-CE97-0003 EQUIPPS. AI tools were used to proofread the manuscript.

\fi

\ifnum 2=\theShow\else
    \bibliography{refs}
\fi


\ifnum 1<\theShow
\appendix
\onecolumngrid

\section{Explicit expression of multimode light states}
In this supplemental material (SM), we detail the properties of the creation and annihilation operators and provide more explicit information on the types of states considered throughout the main text.

We set $\vac$ to be the vacuum state and introduce the annihilation operators $a_i(\lambda)$ (for $i = 0, \dots, n-1$). We assume the usual bosonic commutation relations:
\begin{align}
    [\hat a_i(\lambda_1), \hat a_j(\lambda_2)] &= 0, &
    [\hat a_i^\dagger(\lambda_1), \hat a_j^\dagger(\lambda_2)] &= 0, &
    [\hat a_i(\lambda_1), \hat a_j^\dagger(\lambda_2)] &= \delta(\lambda_1 - \lambda_2)\delta_{i,j}.
\end{align}
As remarked in the main text, the parameter $\lambda$ can encompass both discrete and continuous degrees of freedom. Accordingly, the Dirac delta functions should be understood as encompassing both discrete Kronecker deltas and continuous Dirac deltas. Similarly, integrals should be interpreted as incorporating both discrete sums and continuous integrals. 

A state $\ket{\psi}$ with $k$ photons in the first spatial mode and $l$ photons in the second is characterized by its wave function $F(\lambda_1, \dots, \lambda_k, \mu_1, \dots, \mu_l)$ via
\begin{equation}\label{eq: state k and l photons}
    \ket{\psi} = \int \dd\lambda_1 \cdots \dd\lambda_k \dd\mu_1 \cdots \dd\mu_l \,
    F(\lambda_1, \dots, \lambda_k, \mu_1, \dots, \mu_l) 
    \hat a_1^\dagger(\lambda_1) \cdots \hat a_1^\dagger(\lambda_k) 
    \hat a_2^\dagger(\mu_1) \cdots \hat a_2^\dagger(\mu_l) \vac,
\end{equation}
where the function $F$ can be assumed to be symmetric in its first $k$ arguments and symmetric in its last $l$ arguments. Explicitly, this means that for any $\sigma \in \mathcal S_k$ and $\tau \in \mathcal S_l$ (with $\mathcal S_m$ denoting the symmetric group on $m$ elements), we have
\begin{equation}
    F(\lambda_{\sigma(1)}, \dots, \lambda_{\sigma(k)}, \mu_{\tau(1)}, \dots, \mu_{\tau(l)}) 
    = F(\lambda_1, \dots, \lambda_k, \mu_1, \dots, \mu_l).
\end{equation}
However, note that no symmetry can be assumed between the $\lambda_i$ and $\mu_j$ variables.

The normalization condition for $F$ is obtained by computing $\braket{\psi}$ using the commutation relations and the fact that $\hat a_i(\lambda)\vac = 0$ (which also implies $\bra{\text{vac}} \hat a^\dagger_i(\lambda) = 0$), yielding
\begin{equation}
    \braket{\psi} = k! \, l! \int \dd\lambda_1 \cdots \dd\lambda_k \dd\mu_1 \cdots \dd\mu_l \, 
    \abs{F(\lambda_1, \dots, \lambda_k, \mu_1, \dots, \mu_l)}^2.
\end{equation}
If $F$ is normalized such that the integral of its modulus squared is $1$, then the state must be normalized by a factor of $1/\sqrt{k! \, l!}$ to yield a properly normalized quantum state. This is reminiscent of the usual normalization factor when defining number states in the simple Fock space with one creation operator: $\ket{n} = \frac{(\hat a^\dagger)^n}{\sqrt{n!}} \vac$.

A pure state with an indefinite number of photons is defined as a coherent superposition of states $\ket{\psi_{k,l}}$ of equation~\eqref{eq: state k and l photons}:
\begin{equation}
    \ket{\psi} = \sum_{k,l=0}^\infty \alpha_{k,l} \ket{\psi_{k,l}},
\end{equation}
for arbitrary complex coefficients $\alpha_{k,l}$ such that $\sum_{k,l=0}^\infty \abs{\alpha_{k,l}}^2 = 1$. The extension of all these expressions to the case of $n > 2$ spatial modes is straightforward, albeit resulting in long and cumbersome expressions with a similar structure.

\section{Discussion about symmetry operator \texorpdfstring{$\hat S$}{S}}
In this section, we discuss the role and interpretation of the symmetry operator $\hat S$, introduced in the main text. It is a linear optical operator whose action on the creation operators is given by
\begin{align}
    \hat S\hat a_1^\dagger(\lambda)\hat S^\dagger &= \hat a_2^\dagger(\lambda), & \hat S\hat a_2^\dagger(\lambda)\hat S^\dagger &= \hat a_1^\dagger(\lambda).
\end{align}
Assuming that the vacuum state is invariant under $\hat S$ ({\it i.e.}, $\hat S\vac = \vac$), one can verify that these properties completely characterize the action of $\hat S$ on all quantum states. As previously mentioned, $\hat S$ has the simple physical interpretation of exchanging the two spatial modes. Applying the definition of $\hat S$ twice yields $\hat S^2 = \1$, implying that $\hat S$ is Hermitian. Furthermore, the property $\hat S^2 = \1$ implies that $\hat S$ is a linear symmetry. A basic result from linear algebra tells us that any state can be decomposed into a symmetric and an antisymmetric part
\begin{equation}
    \ket{\psi} = \ket{\psi_s} + \ket{\psi_a} = \frac{\ket{\psi} + \hat S\ket{\psi}}{2} + \frac{\ket{\psi} - \hat S\ket{\psi}}{2},
\end{equation}
with $\hat S\ket{\psi_s} = \ket{\psi_s}$ and $\hat S\ket{\psi_a} = -\ket{\psi_a}$. We can also define the action of $\hat S$ on any Hermitian operator $\hat H$ by $\hat S : \hat H \mapsto \hat S\hat H\hat S$. Similarly, we verify that $\hat S^2 = \1$ when acting on Hermitian operators, confirming that it also acts as a linear symmetry. As such, a similar decomposition of $\hat H$ into symmetric and antisymmetric parts exists:
\begin{equation}
    \hat H = \hat H_s + \hat H_a = \frac{\hat H + \hat S\hat H\hat S}{2} + \frac{\hat H - \hat S\hat H\hat S}{2},
\end{equation}
with $\hat S\hat H_s\hat S = \hat H_s$ and $\hat S\hat H_a\hat S = -\hat H_a$.

We now explain how one can obtain a quantitative measure of the symmetry of any state $\ket{\psi}$ using the operator $\hat S$. Starting with the state $\ket{\psi}$, we decompose it as before: $\ket{\psi} = \ket{\psi_s} + \ket{\psi_a}$. Since these components are not normalized, a meaningful measure of symmetry can be obtained by comparing their relative norms. We thus compute:
\begin{subequations}
    \begin{align}
        \braket{\psi_s} &= \frac{1}{4}(\bra{\psi} + \bra{\psi}\hat S)(\ket{\psi} + \hat S\ket{\psi}), \\
        &= \frac{1}{4}(\braket{\psi} + 2\bra{\psi}\hat S\ket{\psi} + \bra{\psi}\hat S^2\ket{\psi}), \\
        &= \frac{1}{2}(1 + \bra{\psi}\hat S\ket{\psi}),
    \end{align}
\end{subequations}
and similarly, $\braket{\psi_a} = \frac{1}{2}(1 - \bra{\psi}\hat S\ket{\psi})$. If we now define the normalized states $\ket*{\tilde{\psi}_s}$ and $\ket*{\tilde{\psi}_a}$, we can write:
\begin{equation}
    \ket{\psi} = \alpha\ket*{\tilde{\psi}_s} + \beta\ket*{\tilde{\psi}_a},
\end{equation}
with
\begin{align}
    \alpha^2 = \frac{1 + \bra{\psi}\hat S\ket{\psi}}{2}, &\quad \beta^2 = \frac{1 - \bra{\psi}\hat S\ket{\psi}}{2}.
\end{align}
A natural measure of the symmetry is then given by
\begin{equation}
    \alpha^2 - \beta^2 = \bra{\psi}\hat S\ket{\psi}.
\end{equation}
This formula can also be interpreted as the overlap between the original state $\ket{\psi}$ and its swapped version $\hat S\ket{\psi}$. If $\ket{\psi}$ is nearly symmetric, the expectation value approaches $1$; if it is nearly antisymmetric, it approaches $-1$. A simple application of the Cauchy-Schwarz inequality yields:
\begin{equation}
    -1 \leq \bra{\psi}\hat S\ket{\psi} \leq 1.
\end{equation}
We note that the value $\bra{\psi}\hat S\ket{\psi} = 0$ corresponds to an anionic symmetry.

\section{Matrices and linear optical operations}
In this section of the SM, we review, for completeness, the properties of linear optical operations and their matrix representations. A linear optical operation, or mode transformation, that does not mix the internal degree of freedom $\lambda$ is described via the action of unitary matrices. If we consider $M\in\mathcal M_n(\C)$, any $n\times n$ complex matrix, its action on orthogonal modes $\hat a_0^\dagger(\lambda),\dots, \hat a_{n-1}^\dagger(\lambda)$ is given by 
\begin{align}
    M: \hat a_j^\dagger(\lambda)\mapsto \hat b_j^\dagger(\lambda)=\sum_{k=0}^{n-1} M_{kj}\hat a_k^\dagger(\lambda), && \hat a_j(\lambda)\mapsto \hat b_j(\lambda)=\sum_{k=0}^{n-1} M_{kj}^*\hat a_k(\lambda).
\end{align}
Note the choice of index ordering in the matrix $M$. Choosing $M_{jk}$ instead would be equally valid but would yield slightly less appealing formulas. If the $a$ modes are orthogonal ($[\hat a_j(\lambda),\hat a_k^\dagger(\mu)]=\delta(\lambda-\mu)\delta_{jk}$), then it is a valid mode transformation if and only if the $b$ modes are also orthogonal. Indeed,
\begin{subequations}
    \begin{align}
        [\hat b_j(\lambda),\hat b_k^\dagger(\mu)] &= \sum_{l,m=0}^{n-1} M^*_{lj}M_{mk} [\hat a_l(\lambda),\hat a_m^\dagger(\mu)] \\
        &= \sum_{l,m=0}^{n-1} M^*_{lj}M_{mk} \delta(\lambda-\mu)\delta_{lm} \\
        &= \sum_{l=0}^{n-1} M^*_{lj}M_{lk}\delta(\lambda-\mu) = (M^\dagger M)_{jk}\delta(\lambda-\mu),
    \end{align}
\end{subequations}
so the $b$ modes are orthogonal if and only if $M^\dagger M=\1$, {\it i.e.}, as stated above, $M$ is a unitary matrix. From now on, we will denote by $\hat M$ the operator that acts on the Hilbert space via the corresponding mode transformation. Since it is a proper mode transformation, it is automatically unitary (for example, it preserves the norm of states). It is characterized by the relations
\begin{align}
    \hat M\hat a_j^\dagger(\lambda) \hat M^\dagger=\hat b_j^\dagger(\lambda)=\sum_{k=0}^{n-1} M_{kj}\hat a_k^\dagger(\lambda), && \hat M\ket{0}=\ket{0}.
\end{align}
Now, if $M$ and $N$ are two unitary matrices, we can compute the action of $\hat M\hat N$:
\begin{equation}
    \hat M\hat N \hat a_j^\dagger(\lambda)\hat N^\dagger \hat M^\dagger=\sum_{k=0}^{n-1}N_{kj}\hat M\hat a_k^\dagger(\lambda)\hat M^\dagger=\sum_{k,l=0}^{n-1} N_{kj}M_{lk}\hat a_l^\dagger(\lambda)=\sum_{l=0}^{n-1} (MN)_{lj}\hat a_l^\dagger(\lambda).
\end{equation}
Thus, the composition of $\hat M$ and $\hat N$ corresponds exactly to the transformation induced by the matrix product $MN$, or more compactly, $\hat M\hat N=\widehat{MN}$. In the language of group theory, this defines a group homomorphism from $U_n(\C)$ to the group of unitary operators on the $n$-mode Fock space.

For $n=2$, we have simple matrix representations for the symmetry operator $\hat S$ and the beam splitter (BS) $\hat U$. Following our definitions and conventions, we get
\begin{align}
    S=\begin{pmatrix} 0&1\\1&0\end{pmatrix}, && U=\frac{1}{\sqrt{2}}\begin{pmatrix} 1&1\\1&-1\end{pmatrix},
\end{align}
which we recognize, respectively, as the Pauli $X$ matrix and the Hadamard matrix.

\section{Discussion on the BS implementation}
In the Letter, we have considered a BS with a specific choice for its action, which, as discussed in the previous section of the SM, can be described by the Hadamard matrix
\begin{equation}
    \frac{1}{\sqrt{2}}\begin{pmatrix}
        1 & 1\\
        1 & -1
    \end{pmatrix},
\end{equation}
resulting in an inherent asymmetry between the two spatial modes. This explains why our formulas involving the parity operator $\hat \Pi_1$ exhibit an explicit asymmetry in the roles played by the spatial modes. In this section, we discuss alternative choices for the implementation of the BS and their consequences. 

There are many different representations of a balanced BS, and the choice depends on the physical implementation. The most general parametrization of a balanced BS is given by the matrix
\begin{equation}
    \frac{e^{i\tau}}{\sqrt{2}}\begin{pmatrix}
        e^{i\theta} & e^{-i\phi}\\
        e^{i\phi} & -e^{-i\theta}
    \end{pmatrix},
\end{equation}
for $\theta,\phi,\tau \in [-\pi,\pi]$, where the condition that the BS is balanced is ensured by all matrix elements having equal modulus. For common parameter choices, we obtain frequently used BS matrices such as:
\begin{align}
    \frac{1}{\sqrt{2}} \begin{pmatrix}
        1 & 1\\
        1 & -1
    \end{pmatrix}, &&
    \frac{1}{\sqrt{2}} \begin{pmatrix}
        1 & 1\\
        -1 & 1
    \end{pmatrix}, &&
    \frac{1}{\sqrt{2}} \begin{pmatrix}
        1 & i\\
        i & 1
    \end{pmatrix}.
\end{align}

We now address the following question: if we compute $\mathbb P[n_1\equiv 0\pmod{2}]$ using a different BS implementation, which symmetry $\hat{\mathfrak S}$ of the state are we effectively measuring? Let us assume that the operators $\hat{U}$ act via the general matrix
\begin{equation}
    \frac{e^{i\tau}}{\sqrt{2}}\begin{pmatrix}
        e^{i\theta} & e^{-i\phi}\\
        e^{i\phi} & -e^{-i\theta}
    \end{pmatrix}
\end{equation}
as written above. Following the same reasoning as in the main text, the relevant symmetry operator is
\begin{equation}
    \hat{\mathfrak S} = \hat{U}^\dagger \hat{\Pi}_1 \hat{U},
\end{equation}
which corresponds to the matrix product
\begin{equation}
    \left[\frac{e^{i\tau}}{\sqrt{2}}\begin{pmatrix}
        e^{i\theta} & e^{-i\phi}\\
        e^{i\phi} & -e^{-i\theta}
    \end{pmatrix}\right]^\dagger 
    \begin{pmatrix}
        1 & 0\\
        0 & -1
    \end{pmatrix}
    \frac{e^{i\tau}}{\sqrt{2}}\begin{pmatrix}
        e^{i\theta} & e^{-i\phi}\\
        e^{i\phi} & -e^{-i\theta}
    \end{pmatrix}
    = \begin{pmatrix}
        0 & e^{-i(\theta+\phi)}\\ 
        e^{i(\theta+\phi)} & 0
    \end{pmatrix}.
\end{equation}
This implies that the action of $\hat{\mathfrak S}$ is given by
\begin{align}
    \hat{\mathfrak S} \hat{a}_1^\dagger(\lambda) \hat{\mathfrak S}^\dagger = e^{i(\theta+\phi)} \hat{a}_2^\dagger(\lambda), &&
    \hat{\mathfrak S} \hat{a}_2^\dagger(\lambda) \hat{\mathfrak S}^\dagger = e^{-i(\theta+\phi)} \hat{a}_1^\dagger(\lambda).
\end{align}

From these expressions, we can verify that $\hat{\mathfrak S}$ is Hermitian and unitary, satisfying $\hat{\mathfrak S}^2 = \1$. Thus, it defines a valid symmetry: by computing $\mathbb P[n_1\equiv 0\pmod{2}]$, we are probing the symmetry of the state $\ket{\psi}$ with respect to this new symmetry operator $\hat{\mathfrak S}$. We now observe that if we redefine the phases of the creation operators as
\begin{align}
    \hat{b}_1^\dagger(\lambda) = \hat{a}_1^\dagger(\lambda), &&
    \hat{b}_2^\dagger(\lambda) = e^{i(\theta+\phi)} \hat{a}_2^\dagger(\lambda),
\end{align}
then the action of $\hat{\mathfrak S}$ on the new operators $\hat{b}_i^\dagger(\lambda)$ becomes
\begin{align}
    \hat{b}_1^\dagger(\lambda) \mapsto \hat{b}_2^\dagger(\lambda), &&
    \hat{b}_2^\dagger(\lambda) \mapsto \hat{b}_1^\dagger(\lambda),
\end{align}
which corresponds to the standard symmetry relations defined by $\hat{S}$. This demonstrates that the only difference lies in the phase conventions for each spatial mode.

Finally, we note that if $\ket{\psi}$ is a state with the same number of photons in each mode (or a coherent superposition of such states), then the action of $\hat{\mathfrak S}$ is equivalent to that of $\hat{S}$. Indeed, the phase factors $e^{i(\theta+\phi)}$ and $e^{-i(\theta+\phi)}$ appear in equal numbers and cancel out. This is particularly relevant for the standard HOM experiment, where two single photons are input into different arms. Thus, we recover the well-known result that the specific implementation of a balanced BS does not affect the coincidence probability in HOM interferometry. This explains why such discussions are often omitted in that context. However, in more general settings, the implementation of the BS can indeed play a significant role.

\section{Diagonalisation via Fourier matrices}
In this section of the SM, we provide a detailed verification of a simple and well-known formula for the diagonalization of a permutation matrix. The primary aim is to present an explicit derivation of the result. In the main text, we introduced the operator $\hat P$, characterized by the matrix
\begin{equation}
    P = \begin{pmatrix} 
        0 & 1 & & \\
          & \ddots & \ddots & \\
          &        & \ddots & 1 \\
        1 &        &        & 0 
    \end{pmatrix},
\end{equation}
which consists of $n$ ones and zeros elsewhere. The ones are placed along the first upper diagonal ($n-1$ entries), with a single 1 in the lower-left corner. Interpreting $P$ as a discrete translation operator with periodic boundary conditions, it is natural to seek its eigenvectors in the form of (discrete) plane waves. Define $e_j = (1, \omega^j, \omega^{2j}, \dots, \omega^{(n-1)j})^T$, where $\omega = e^{2\pi i / n}$ is a primitive $n$th root of unity. A straightforward computation shows that
\begin{equation}
    P e_j = P 
    \begin{pmatrix} 
        1 \\ \omega^j \\ \vdots \\ \omega^{(n-1)j} 
    \end{pmatrix} 
    = \begin{pmatrix} 
        \omega^j \\ \vdots \\ \omega^{(n-1)j} \\ 1 
    \end{pmatrix} 
    = \omega^j 
    \begin{pmatrix} 
        1 \\ \vdots \\ \omega^{(n-2)j} \\ \omega^{(n-1)j} 
    \end{pmatrix} 
    = \omega^j e_j.
\end{equation}
Thus, the vectors $(e_0, \dots, e_{n-1})$ form an eigenbasis of $P$. Normalizing by $1/\sqrt{n}$ and defining the matrix of eigenvectors as $U = \frac{1}{\sqrt{n}} \begin{pmatrix} e_0 & e_1 & \cdots & e_{n-1} \end{pmatrix}$, we obtain the diagonalization
\begin{equation}
    P = U D U^{-1},
\end{equation}
where $D = \operatorname{diag}(1, \omega, \dots, \omega^{n-1})$. Observing that the entries of $U$ are given by $U_{kl} = \frac{1}{\sqrt{n}} \omega^{kl}$ (with indices ranging from $0$ to $n-1$), we identify $U$ as the discrete Fourier transform (DFT) matrix. It is well known (or can be readily verified) that such a matrix is unitary. We thus obtain
\begin{equation}
    P = U D U^\dagger.
\end{equation}
Taking the complex conjugate (not the Hermitian conjugate) and noting that $U^* = U^\dagger$ and $D^* = D^\dagger$, we also find
\begin{equation}
    P = U^\dagger D^\dagger U.
\end{equation}
Finally, from the results established in the SM on linear optical operations, if we associate to these matrices the corresponding operators $\hat U$, $\hat P$, and $\hat D$, we obtain the operator identities
\begin{align}
    \hat P = \hat U \hat D \hat U^\dagger, \qquad \hat P = \hat U^\dagger \hat D^\dagger \hat U.
\end{align}

\section{Fourier interferometer photon number formula}
In this section, we formally prove the formula expressing the probability of measuring a certain photon number distribution at the output of a DFT interferometer, in terms of the expectation value of powers of $\hat P$ evaluated on the initial state
\begin{equation}
    \mathbb P\left[\sum_{k=0}^{n-1} k m_k\equiv0\pmod{n}\right]=\frac{1}{n}\sum_{l=0}^{n-1} \bra{\psi}P^l\ket{\psi}.
\end{equation}
To do so, we first show how, in the absence of the interferometer, expectation values of $\hat D$ relate to the photon number distribution. The situation is as follows: we consider a general quantum state $\ket{\psi}$ distributed over $n$ modes, with an arbitrary photon number distribution in each mode and an arbitrary distribution over internal degrees of freedom. We assume that photon numbers can be measured in each mode, irrespective of the internal degree of freedom. To build intuition, we begin with simple pure states and gradually generalize.

Temporarily setting aside the dependence of the operator on $\lambda$, we observe that for integers $m_0,\dots,m_{n-1}$,
\begin{subequations}
    \begin{align}
        \hat D (\hat a_0^\dagger)^{m_0}\cdots (\hat a_{n-1}^\dagger)^{m_{n-1}}\hat D^\dagger &= (\hat D\hat a_0^\dagger\hat D^\dagger)^{m_0}\cdots (\hat D\hat a_{n-1}^\dagger\hat D^\dagger)^{m_{n-1}} \\
        &= (\omega^0\hat a_0^\dagger)^{m_0}\cdots (\omega^{n-1}\hat a_{n-1}^\dagger)^{m_{n-1}} \\
        &= \omega^{\sum_{k=0}^{n-1}km_k}(\hat a_0^\dagger)^{m_0}\cdots (\hat a_{n-1}^\dagger)^{m_{n-1}}.
    \end{align}
\end{subequations}
This implies that if $\ket{\psi} = \frac{1}{\sqrt{m_0!\cdots m_{n-1}!}}(\hat a_0^\dagger)^{m_0}\cdots (\hat a_{n-1}^\dagger)^{m_{n-1}}\vac$ is a Fock state with a definite number of photons in each spatial mode, then
\begin{subequations}
    \begin{align}
        \hat D\ket{\psi} &= \frac{1}{\sqrt{m_0!\cdots m_{n-1}!}}\hat D(\hat a_0^\dagger)^{m_0}\cdots (\hat a_{n-1}^\dagger)^{m_{n-1}}\hat D^\dagger \hat D\vac \\
        &= \frac{\omega^{\sum_{k=0}^{n-1}km_k}}{\sqrt{m_0!\cdots m_{n-1}!}}(\hat a_0^\dagger)^{m_0}\cdots (\hat a_{n-1}^\dagger)^{m_{n-1}}\vac \\
        &= \omega^{\sum_{k=0}^{n-1}km_k}\ket{\psi},
    \end{align}
\end{subequations}
so that $\expval*{\hat D} := \bra{\psi}\hat D\ket{\psi} = \omega^{\sum_{k=0}^{n-1}km_k}$.

The extension to states including internal degrees of freedom is similar, though more cumbersome to write explicitly. A general state $\ket{\psi}$ with $m_j$ photons in mode $j$ and an arbitrary wave function is written as
\begin{equation}\label{eq: equation n mode tf state with fixed nbr of photons}
    \ket{\psi} = \int \dd\{\lambda_{jk}\} F(\{\lambda_{jk}\})\prod_{j,k} \hat a_j(\lambda_{jk})^\dagger\vac,
\end{equation}
where the integration is over all variables $\lambda_{jk}$ with $j=0,\dots,n-1$ and $k=1,\dots, m_j$. The function $F$ depends on all these variables. Since the action of $\hat D$ is independent of the $\lambda_{jk}$, only the number of creation operators in each spatial mode matters. Therefore, we again have $\hat D\ket{\psi} = \omega^{\sum_{k=0}^{n-1}km_k}\ket{\psi}$, and consequently $\expval*{\hat D} = \omega^{\sum_{k=0}^{n-1}km_k}$.

For such a state $\ket{\psi}$ with $m_k$ photons in each mode, the expectation value of powers of $\hat D$ is readily computed. Since $\hat D^l\ket{\psi} = \omega^{l\sum_{k=0}^{n-1}km_k}\ket{\psi}$, it follows that $\expval*{\hat D^l} = \omega^{l\sum_{k=0}^{n-1}km_k}$. Summing over $l=0,\dots,n-1$ and using the geometric series formula, we obtain
\begin{equation}
    \frac{1}{n}\sum_{l=0}^{n-1}\expval*{\hat D^l} = \frac{1}{n}\sum_{l=0}^{n-1} \left(\omega^{\sum_{k=0}^{n-1}km_k}\right)^l = \left\{\begin{array}{cc}
         1 & \text{if } \sum_{k=0}^{n-1}km_k \equiv 0 \pmod{n}, \\
         0 & \text{otherwise}.
    \end{array}\right.
\end{equation}

The most general pure state over $n$ time-frequency modes can be written as
\begin{equation}
    \ket{\psi} = \sum_{m_0,\dots,m_{n-1}=0}^\infty \alpha_{\vec{m}}\ket{\psi_{\vec{m}}},
\end{equation}
where $\ket{\psi_{\vec{m}}}$ is a state with $m_j$ photons in mode $j$, of the form given in Eq.~\eqref{eq: equation n mode tf state with fixed nbr of photons}, and $\alpha_{\vec{m}} \in \mathbb{C}$ with $\sum_{\vec{m}}|\alpha_{\vec{m}}|^2 = 1$. As states with different photon numbers are orthogonal, $\braket{\psi_{\vec m}}{\psi_{\vec m'}} = \delta_{\vec m, \vec m'}$, and the action of $\hat D$ preserves photon number, we compute
\begin{subequations}
    \begin{align}
        \frac{1}{n}\sum_{l=0}^{n-1}\expval{\hat D^l}_{\ket{\psi}} &= \frac{1}{n}\sum_{l=0}^{n-1} \sum_{\vec m,\vec m'} \alpha_{\vec m}^*\alpha_{\vec m'}\bra{\psi_{\vec m}}\hat D^l\ket{\psi_{\vec m'}} \\
        &= \frac{1}{n}\sum_{l=0}^{n-1} \sum_{\vec m} |\alpha_{\vec m}|^2 \omega^{l\sum_{k=0}^{n-1}km_k} \\
        &= \sum_{\vec m} |\alpha_{\vec m}|^2 \delta_{\sum_{k=0}^{n-1}km_k \equiv 0\pmod{n}} \\
        &= \sum_{\substack{\vec m \\ \sum_{k=0}^{n-1}km_k \equiv 0\pmod{n}}} |\alpha_{\vec m}|^2 = \mathbb{P}\left[\sum_{k=0}^{n-1}km_k \equiv 0 \pmod{n} \right].
    \end{align}
\end{subequations}

We now insert the interferometer $\hat U$ between the state and number-resolving photon detection. Based on the result above and using the identity $\hat P = \hat U^\dagger \hat D^\dagger \hat U$, the probability that the measured photon numbers satisfy $\sum_{k=0}^{n-1} k m_k \equiv 0\pmod{n}$ is
\begin{subequations}
    \begin{align}
        \mathbb{P}\left[\sum_{k=0}^{n-1}km_k \equiv 0\pmod{n}\right] &= \frac{1}{n}\sum_{l=0}^{n-1} \expval{\hat D^l}_{\hat U\ket{\psi}} = \frac{1}{n}\sum_{l=0}^{n-1} \bra{\psi} \hat U^\dagger \hat D^l \hat U \ket{\psi} \\
        &= \frac{1}{n}\sum_{l=0}^{n-1} \bra{\psi} \hat U^\dagger (\hat D^\dagger)^{-l} \hat U \ket{\psi} = \frac{1}{n}\sum_{l=0}^{n-1} \bra{\psi} \hat P^{-l} \ket{\psi} \\
        &= \frac{1}{n}\sum_{l=0}^{n-1} \expval{\hat P^l}_{\ket{\psi}},
    \end{align}
\end{subequations}
where we used $\hat P^n = \1$ and the change of index $l \to n - l$. Thus, the probability of observing a photon number distribution satisfying the modular constraint is determined by the symmetry of the incoming state under the cyclic permutation $\sigma$. Since $\hat D$ diagonalizes $\hat P$, the eigenvalues of $\hat P$ are exactly $1, \omega, \dots, \omega^{n-1}$. If $\ket{\psi}$ is an eigenstate of $\hat P$ with eigenvalue $\theta = \omega^k$, then
\begin{equation}
    \mathbb{P}\left[\sum_{k=0}^{n-1} k m_k \equiv 0\pmod{n}\right] = \frac{1}{n} \sum_{l=0}^{n-1} \theta^l = \left\{
    \begin{array}{cc}
        1 & \text{if } \theta = 1, \\
        0 & \text{otherwise}.
    \end{array}
    \right.
\end{equation}
Therefore, eigenstates of $\hat P$ determine the extremal values of the probability. Furthermore, it means that the operator 
\begin{equation}
    \hat \Pi=\frac{1}{n}\sum_{l=0}^{n-1} \hat P^l,
\end{equation}
can be understood as the projector onto the subspace of eigenvectors $\ket{\psi}$ of $\hat P$ such that $\hat P\ket{\psi}=\ket{\psi}$.

\section{Permutation beyond cycles}
In this section, we explore alternative choices for the permutation operator $\hat{P}$ and justify why the choice made in the main text is the most compelling. Instead of a cyclic permutation, we could consider any permutation of the modes. We thus denote by $\sigma \in \mathfrak{S}_n$ a generic permutation of the set $\{0,\dots,n-1\}$. We define $\hat{P}_\sigma$ as the mode transformation operator satisfying
\begin{equation}
    \hat{P}_\sigma \, \hat{a}_j^\dagger(\lambda) \, \hat{P}_\sigma^\dagger = \hat{a}_{\sigma(j)}^\dagger(\lambda),
\end{equation}
and denote by $P_\sigma$ the associated matrix. As in the case presented in the main text, $P_\sigma$ is a permutation matrix. It is a well-known result in the theory of permutations that any $\sigma$ can be decomposed into a product of disjoint cycles
\begin{equation}
    \sigma = c_1 \circ c_2 \circ \cdots \circ c_k,
\end{equation}
where each $c_j$ is a cycle, {\it i.e.}, a permutation of the form $\alpha_1 \mapsto \alpha_2 \mapsto \cdots \mapsto \alpha_\ell \mapsto \alpha_1$. This implies that, up to a relabeling of the spatial modes, the matrix $P_\sigma$ can be brought into a block-diagonal form:
\begin{equation}
    P_\sigma = \begin{pmatrix}
        \boxed{P_1} & 0 & \cdots & 0 \\
        0 & \boxed{P_2} & \cdots & 0 \\
        \vdots & \vdots & \ddots & \vdots \\
        0 & 0 & \cdots & \boxed{P_k}
    \end{pmatrix},
\end{equation}
where each $P_j$ is a permutation matrix corresponding to a single cycle, of the same form as the matrix $P$ considered in the main text:
\begin{equation}
    P_j = \begin{pmatrix} 
        0 & 1 & & \\
        & \ddots & \ddots & \\
        & & \ddots & 1 \\
        1 & & & 0 
    \end{pmatrix}.
\end{equation}
Each block $P_j$ can be diagonalized using a DFT matrix of the corresponding size, implying that $P_\sigma$ itself is diagonalized by $U_1 \oplus \cdots \oplus U_k$. Translating this into the operator language, the interferometer adapted to the transformation $\hat{P}_\sigma$ is composed of $k$ independent smaller DFT interferometers, as depicted in Fig.~\ref{fig:many-fourier}. Such an interferometer does not mix the different groups of modes. Consequently, for metrological purposes, the entanglement present in the state $\ket{\psi}$ between different mode groups cannot be exploited to enhance estimation precision. This observation justifies why such architectures were not considered in the main text.

\begin{figure}
    \centering
    \scalebox{1.4}{\begin{tikzpicture}
	\begin{pgfonlayer}{nodelayer}
		\node [style=none] (0) at (0.75, 2) {$\hat U_1$};
		\node [style=none] (1) at (-0.25, 1) {};
		\node [style=none] (2) at (1.75, 1) {};
		\node [style=none] (3) at (-0.25, 3) {};
		\node [style=none] (4) at (1.75, 3) {};
		\node [style=none] (5) at (-0.25, 2.5) {};
		\node [style=none] (6) at (-0.25, 1.5) {};
		\node [style=none] (7) at (-0.25, -2) {};
		\node [style=none] (8) at (-0.25, -1) {};
		\node [style=none] (9) at (-1.75, 2.5) {};
		\node [style=none] (10) at (-1.75, 1.5) {};
		\node [style=none] (11) at (-1.75, -1) {};
		\node [style=none] (12) at (-1.75, -2) {};
		\node [style=none] (13) at (-1, 0.5) {$\vdots$};
		\node [style=none] (14) at (2.75, 2.5) {};
		\node [style=none] (15) at (2.75, 1.5) {};
		\node [style=none] (16) at (2.75, -2) {};
		\node [style=none] (17) at (2.75, -1) {};
		\node [style=none] (18) at (1.75, 2.5) {};
		\node [style=none] (19) at (1.75, 1.5) {};
		\node [style=none] (20) at (1.75, -1) {};
		\node [style=none] (21) at (1.75, -2) {};
		\node [style=none] (22) at (2.25, 0.5) {$\vdots$};
		\node [style=none] (23) at (3.5, 2.75) {};
		\node [style=none] (24) at (3.5, 2.25) {};
		\node [style=none] (25) at (4, 2.5) {};
		\node [style=none] (26) at (4.5, 2.5) {};
		\node [style=none] (27) at (5.25, 2.5) {$m_0$};
		\node [style=none] (28) at (3.5, 1.75) {};
		\node [style=none] (29) at (3.5, 1.25) {};
		\node [style=none] (30) at (4, 1.5) {};
		\node [style=none] (31) at (4.5, 1.5) {};
		\node [style=none] (32) at (5.25, 1.5) {$m_l$};
		\node [style=none] (33) at (5.5, -1) {$m_j$};
		\node [style=none] (34) at (5.5, -2) {$m_{n-1}$};
		\node [style=none] (35) at (-6.25, 0.25) {$\ket{\psi}$};
		\node [style=none] (36) at (-5, 0.25) {\scalebox{1}[8]{$\{$}};
		\node [style=none] (37) at (-2.75, 0.25) {$e^{i\kappa\hat H}$};
		\node [style=none] (38) at (-3.5, -2.75) {};
		\node [style=none] (39) at (-2, -2.75) {};
		\node [style=none] (40) at (-3.5, 3.25) {};
		\node [style=none] (41) at (-2, 3.25) {};
		\node [style=none] (42) at (-3.75, 2.5) {};
		\node [style=none] (43) at (-3.75, 1.5) {};
		\node [style=none] (44) at (-3.75, -2) {};
		\node [style=none] (45) at (-3.75, -1) {};
		\node [style=none] (46) at (-4.75, 2.5) {};
		\node [style=none] (47) at (-4.75, 1.5) {};
		\node [style=none] (48) at (-4.75, -1) {};
		\node [style=none] (49) at (-4.75, -2) {};
		\node [style=none] (50) at (-4.25, 0.5) {$\vdots$};
		\node [style=none] (51) at (3.5, -0.75) {};
		\node [style=none] (52) at (3.5, -1.25) {};
		\node [style=none] (53) at (4, -1) {};
		\node [style=none] (54) at (4.5, -1) {};
		\node [style=none] (55) at (3.5, -1.75) {};
		\node [style=none] (56) at (3.5, -2.25) {};
		\node [style=none] (57) at (4, -2) {};
		\node [style=none] (58) at (4.5, -2) {};
		\node [style=none] (59) at (-3.75, 3) {};
		\node [style=none] (60) at (-1.75, 3) {};
		\node [style=none] (61) at (-1.75, -2.5) {};
		\node [style=none] (62) at (-3.75, -2.5) {};
		\node [style=none] (63) at (-4.5, -0.25) {};
		\node [style=none] (66) at (-4.25, 2.125) {\scalebox{0.6}{$\vdots$}};
		\node [style=none] (67) at (2.25, 2.125) {\scalebox{0.6}{$\vdots$}};
		\node [style=none] (68) at (-1, 2.125) {\scalebox{0.6}{$\vdots$}};
		\node [style=none] (69) at (0.75, -1.5) {$\hat U_k$};
		\node [style=none] (70) at (-0.25, -2.5) {};
		\node [style=none] (71) at (-0.25, -0.5) {};
		\node [style=none] (72) at (1.75, -0.5) {};
		\node [style=none] (73) at (-4.25, -1.375) {\scalebox{0.6}{$\vdots$}};
		\node [style=none] (74) at (2.25, -1.375) {\scalebox{0.6}{$\vdots$}};
		\node [style=none] (75) at (-1, -1.375) {\scalebox{0.6}{$\vdots$}};
		\node [style=none] (76) at (1.75, -2.5) {};
	\end{pgfonlayer}
	\begin{pgfonlayer}{edgelayer}
		\draw [style=Fill grey] (2.center)
			 to (1.center)
			 to (3.center)
			 to (4.center)
			 to cycle;
		\draw (9.center) to (5.center);
		\draw (6.center) to (10.center);
		\draw (11.center) to (8.center);
		\draw (7.center) to (12.center);
		\draw (18.center) to (14.center);
		\draw (15.center) to (19.center);
		\draw (20.center) to (17.center);
		\draw (16.center) to (21.center);
		\draw [style=Fill pink] (24.center)
			 to (23.center)
			 to [in=90, out=0, looseness=0.75] (25.center)
			 to [in=0, out=-90, looseness=0.75] cycle;
		\draw (25.center) to (26.center);
		\draw [style=Fill pink] (29.center)
			 to (28.center)
			 to [in=90, out=0, looseness=0.75] (30.center)
			 to [in=0, out=-90, looseness=0.75] cycle;
		\draw (30.center) to (31.center);
		\draw (60.center)
			 to [bend right=45, looseness=1.25] (41.center)
			 to (40.center)
			 to [bend right=45, looseness=1.25] (59.center)
			 to (62.center)
			 to [bend right=45] (38.center)
			 to (39.center)
			 to [bend right=45] (61.center)
			 to cycle;
		\draw (46.center) to (42.center);
		\draw (43.center) to (47.center);
		\draw (48.center) to (45.center);
		\draw (44.center) to (49.center);
		\draw [style=Fill pink] (52.center)
			 to (51.center)
			 to [in=90, out=0, looseness=0.75] (53.center)
			 to [in=0, out=-90, looseness=0.75] cycle;
		\draw (53.center) to (54.center);
		\draw [style=Fill pink] (56.center)
			 to (55.center)
			 to [in=90, out=0, looseness=0.75] (57.center)
			 to [in=0, out=-90, looseness=0.75] cycle;
		\draw (57.center) to (58.center);
		\draw [style=Fill grey] (76.center)
			 to [in=0, out=180] (70.center)
			 to (71.center)
			 to (72.center)
			 to cycle;
	\end{pgfonlayer}
\end{tikzpicture}}
    \caption{Schematic of the interferometer adapted to access the symmetry associated with a general permutation operator $\hat{P}_\sigma$.}
    \label{fig:many-fourier}
\end{figure}
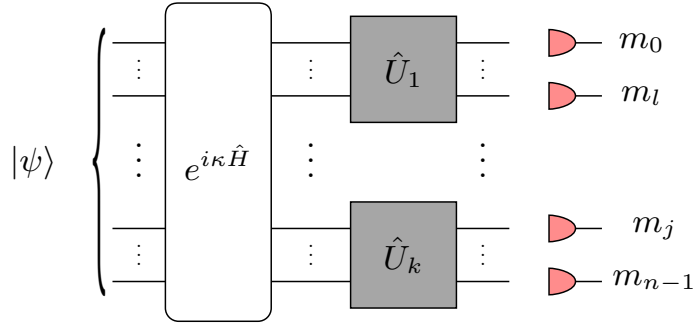

\section{Other results for the linear combination \texorpdfstring{$\sum_{k=0}^{n-1} k m_k$}{sum of k mk}}
In the main text and in the previous section of the SM, we argued, based on the equation
\begin{equation}
    \mathbb{P}\left[\sum_{k=0}^{n-1} k m_k \equiv 0\pmod{n}\right] 
    = \frac{1}{n}\sum_{l=0}^{n-1} \expval{P^l}_{\ket{\psi}} 
    = \expval{\hat \Pi}_{\ket{\psi}},
\end{equation}
that the probability that the measured photon numbers satisfy $\sum_{k=0}^{n-1} k m_k \equiv 0 \pmod{n}$ equals the squared norm of the component of $\ket{\psi}$ in the eigenspace of $\hat P$ associated with the eigenvalue $+1$. In this section, we describe how other values of $\sum_{k=0}^{n-1} k m_k$ provide insight into the norm of the component of $\ket{\psi}$ in the other eigenspaces of $\hat P$. First, observe that for $j \in \{0, \dots, n-1\}$, the operator
\begin{equation}
    \hat \Pi_j = \frac{1}{n} \sum_{l=0}^{n-1} (\omega^{-j} \hat P)^l
\end{equation}
is a projector (it is straightforward to verify that it is Hermitian) onto the eigenspace of $\hat P$ associated with the eigenvalue $\omega^j$. Recall that the eigenvalues of $\hat P$ are precisely $1, \omega, \dots, \omega^{n-1}$. If $\ket{\psi}$ is an eigenstate of $\hat P$ with eigenvalue $\omega^\ell$, we have
\begin{equation}
    \hat \Pi_j \ket{\psi} = \frac{1}{n} \sum_{l=0}^{n-1} (\omega^{-j} \hat P)^l \ket{\psi} 
    = \frac{1}{n} \sum_{l=0}^{n-1} (\omega^{-j} \omega^\ell)^l \ket{\psi}
    = \left\{
    \begin{array}{cl}
         \ket{\psi}, & \text{if } j = \ell, \\
         0, & \text{otherwise}.
    \end{array}
    \right.
\end{equation}

Now consider an initial state $\ket{\psi}$ for which $\ket{\varphi} = \hat U \ket{\psi}$ has exactly $m_k$ photons in output mode $k$. Then,
\begin{subequations}
    \begin{align}
        \expval{\hat \Pi_j}_{\ket{\psi}} 
        &= \frac{1}{n} \sum_{l=0}^{n-1} \omega^{-j l} \bra{\psi} \hat P^l \ket{\psi} 
        = \frac{1}{n} \sum_{l=0}^{n-1} \omega^{-j l} \bra{\psi} \hat U^\dagger \hat U \hat P^l \hat U^\dagger \hat U \ket{\psi} \\
        &= \frac{1}{n} \sum_{l=0}^{n-1} \omega^{-j l} \bra{\varphi} (\hat D^\dagger)^l \ket{\varphi} 
        = \frac{1}{n} \sum_{l=0}^{n-1} \omega^{-j l} \left( \omega^{-\sum_{k=0}^{n-1} k m_k} \right)^l \\
        &= \frac{1}{n} \sum_{l=0}^{n-1} \left( \omega^{-j - \sum_{k=0}^{n-1} k m_k} \right)^l \\
        &= \left\{
        \begin{array}{cl}
            1, & \text{if } \sum_{k=0}^{n-1} k m_k \equiv -j \pmod{n}, \\
            0, & \text{otherwise}.
        \end{array}
        \right.
    \end{align}
\end{subequations}
Following the reasoning presented in a previous part of the SM, this implies that
\begin{equation}
    \mathbb{P}\left[\sum_{k=0}^{n-1} k m_k \equiv -j \pmod{n}\right] 
    = \expval{\hat \Pi_j}_{\ket{\psi}},
\end{equation}
demonstrating that values of $\sum_{k=0}^{n-1} k m_k$ modulo $n$ probe the squared norm of the component of $\ket{\psi}$ in the other eigenspaces of $\hat P$.

\section{FI and small parameter expansion}
In this short section, we show how to compute the Fisher information (FI) in the vicinity of $\kappa = 0$ under certain assumptions on the probabilities. We assume that the experiment has only two outcomes, with probabilities denoted by $p(\kappa)$ and $q(\kappa)$, where $p(0) = 1$. Furthermore, we assume that the probabilities can be expanded near $\kappa = 0$ as
\begin{equation}
    p(\kappa) = 1 - \kappa^2 A + o(\kappa^2), \quad q(\kappa) = \kappa^2 A + o(\kappa^2)
\end{equation}
for some arbitrary non-negative constant $A$. In this case, for $\kappa \neq 0$, the FI is given by
\begin{subequations}
    \begin{align}
        \mathcal{F} &= \frac{1}{p(\kappa)}\left(\frac{\dd p(\kappa)}{\dd\kappa}\right)^2 + \frac{1}{q(\kappa)}\left(\frac{\dd q(\kappa)}{\dd\kappa}\right)^2 \\
        &= \frac{(p'(\kappa))^2}{p(\kappa)(1 - p(\kappa))} \\
        &= \frac{(2\kappa A + o(\kappa^2))^2}{\kappa^2 A + o(\kappa^2))(1 - \kappa^2 A + o(\kappa^2))} \\
        &= \frac{4\kappa^2 A^2 + o(\kappa^2)}{\kappa^2 A + o(\kappa^2)} \\
        &= 4A + o(1)
    \end{align}
\end{subequations}
so that
\begin{equation}
    \lim_{\kappa \to 0} \mathcal{F}(\kappa) = 4A.
\end{equation}
The precision is thus directly proportional to the parameter $A$ defined in the expansion above.

\section{Explicit expressions of the FI}
In this section, we derive the different expressions for the FI presented in the main text. To do so, we utilize the expansion performed in the previous section. We begin by considering a state arriving at the DFT interferometer of the form $e^{i\kappa \hat{H}} \ket{\psi}$ and expand the expression of $\mathbb{P}\left[\sum_{k=0}^{n-1} k m_k \equiv 0 \;[n] \,\middle|\, \kappa \right]$.
\begin{subequations}\label{eq: expansion second order pure}
    \begin{align}
        \mathbb{P}\left[\sum_{k=0}^{n-1} k m_k \equiv 0\;[n] \,\middle|\, \kappa\right] 
        &= \frac{1}{n}\sum_{l=0}^{n-1} \bra{\psi} e^{-i\kappa \hat{H}} \hat{P}_\sigma^l e^{i\kappa \hat{H}} \ket{\psi} \\
        &= \bra{\psi} e^{-i\kappa \hat{H}} \hat{\Pi} e^{i\kappa \hat{H}} \ket{\psi} \\
        &= \expval{ \left(1 - i\kappa \hat{H} - \tfrac{\kappa^2}{2} \hat{H}^2 \right) \hat{\Pi} \left(1 + i\kappa \hat{H} - \tfrac{\kappa^2}{2} \hat{H}^2 \right) } + o(\kappa^2) \\
        &= \expval{ \hat{\Pi} + i\kappa (\hat{\Pi} \hat{H} - \hat{H} \hat{\Pi}) - \tfrac{\kappa^2}{2} \left( \hat{H}^2 \hat{\Pi} + \hat{\Pi} \hat{H}^2 - 2\hat{H} \hat{\Pi} \hat{H} \right)} + o(\kappa^2) \\
        &= \expval{\hat{\Pi}} + i\kappa \left( \expval{\hat{\Pi} \hat{H}} - \expval{\hat{H} \hat{\Pi}} \right) - \tfrac{\kappa^2}{2} \left( \expval{\hat{H}^2 \hat{\Pi}} + \expval{\hat{\Pi} \hat{H}^2} - 2\expval{\hat{H} \hat{\Pi} \hat{H}} \right) + o(\kappa^2).
    \end{align}
\end{subequations}
We want this equation to take the form $p(\kappa)$ or $q(\kappa)$ from the previous section, which requires that $\expval{\hat{\Pi}} = 1$ or $0$. As the computations differ slightly, we treat each case separately.

$\blacktriangleright$ Let's first assume that $\expval{\hat{\Pi}} = 1$. Since $\hat{\Pi}$ is a projector, this implies that $\hat{\Pi} \ket{\psi} = \ket{\psi}$. As such, we can simplify the previous expression using the fact that each occurrence of $\hat{\Pi}$ to the left or right of an expectation value can be removed. We obtain
\begin{equation}
    \mathbb{P}\left[\sum_{k=0}^{n-1} k m_k \equiv 0\;[n] \,\middle|\, \kappa\right] = 1 - \kappa^2 \left( \expval{\hat{H}^2} - \expval{\hat{H} \hat{\Pi} \hat{H}} \right) + o(\kappa^2).
\end{equation}
We observe that the term $\expval*{\hat{H}^2} - \expval*{\hat{H} \hat{\Pi} \hat{H}}$ can be expressed as the variance of two different operators
\begin{equation}\label{eq: second order for sym}
    \expval*{\hat{H}^2} - \expval*{\hat{H} \hat{\Pi} \hat{H}} = \Delta(i[\hat{H}, \hat{\Pi}]) = \frac{1}{n^2} \Delta\left(n \hat{H} - \sum_{l=0}^{n-1} \hat{P}^l \hat{H} \hat{P}^{-l} \right).
\end{equation}
Let us first verify the left-hand identity. The commutator of two Hermitian operators is anti-Hermitian, hence, we use $i[\hat{H}, \hat{\Pi}]$ to obtain a Hermitian operator. Since $\hat{\Pi} \ket{\psi} = \ket{\psi}$, we find:
\begin{equation}
    \expval{i[\hat{H}, \hat{\Pi}]} = i\expval{\hat{H} \hat{\Pi}} - i\expval{\hat{\Pi} \hat{H}} = i\expval{\hat{H}} - i\expval{\hat{H}} = 0.
\end{equation}
Then,
\begin{subequations}
    \begin{align}
        \expval{(i[\hat{H}, \hat{\Pi}])^2} &= -\expval{(\hat{H} \hat{\Pi} - \hat{\Pi} \hat{H})^2} \\
        &= -\expval{ \hat{H} \hat{\Pi} \hat{H} \hat{\Pi} - \hat{H} \hat{\Pi}^2 \hat{H} - \hat{\Pi} \hat{H}^2 \hat{\Pi} + \hat{\Pi} \hat{H} \hat{\Pi} \hat{H} } \\
        &= -\expval{\hat{H} \hat{\Pi} \hat{H}} + \expval{\hat{H} \hat{\Pi} \hat{H}} + \expval{\hat{H}^2} - \expval{\hat{H} \hat{\Pi} \hat{H}} \\
        &= \expval{\hat{H}^2} - \expval{\hat{H} \hat{\Pi} \hat{H}}.
    \end{align}
\end{subequations}
To verify the second expression, we use the identity $\hat{P} \ket{\psi} = \ket{\psi}$, which implies we can remove $\hat{P}$ when it appears on the left or right of an expectation value
\begin{equation}
    \expval{n\hat{H} - \sum_{l=0}^{n-1} \hat{P}^l \hat{H} \hat{P}^{-l}} = n \expval{\hat{H}} - \sum_{l=0}^{n-1} \expval{\hat{H}} = 0.
\end{equation}
Then,
\begin{subequations}
    \begin{align}
        &\expval{\left(n\hat{H} - \sum_{l=0}^{n-1} \hat{P}^l \hat{H} \hat{P}^{-l} \right)^2} \\
        &= n^2 \expval{\hat{H}^2} - 2n \sum_{l=0}^{n-1} \expval{\hat{H} \hat{P}^{-l} \hat{H}} + \sum_{k,l=0}^{n-1} \expval{\hat{H} \hat{P}^{l-k} \hat{H}} \\
        &= n^2 \expval{\hat{H}^2} - 2n \sum_{l=0}^{n-1} \expval{\hat{H} \hat{P}^{-l} \hat{H}} + n \sum_{l=0}^{n-1} \expval{\hat{H} \hat{P}^{l} \hat{H}} \\
        &= n^2 \left( \expval{\hat{H}^2} - \expval{\hat{H} \hat{\Pi} \hat{H}} \right).
    \end{align}
\end{subequations}

$\blacktriangleright$ Now assume $\expval{\hat{\Pi}} = 0$. Since $\hat{\Pi}$ is a projector, this implies $\hat{\Pi} \ket{\psi} = 0$. Thus, any expectation value involving $\hat{\Pi}$ at either end vanishes
\begin{equation}
    \mathbb{P}\left[\sum_{k=0}^{n-1} k m_k \equiv 0\;[n] \,\middle|\, \kappa\right] = \kappa^2 \expval*{\hat{H} \hat{\Pi} \hat{H}} + o(\kappa^2).
\end{equation}
As before, the quadratic term can be expressed as a variance
\begin{equation}\label{eq: second order non sym}
    \expval*{\hat{H} \hat{\Pi} \hat{H}} = \Delta(i[\hat{H}, \hat{\Pi}]).
\end{equation}
Indeed, 
\begin{equation}
    \expval{i[\hat{H}, \hat{\Pi}]} = i\expval{\hat{H} \hat{\Pi}} - i\expval{\hat{\Pi} \hat{H}} = 0,
\end{equation}
and
\begin{subequations}
    \begin{align}
        \expval{(i[\hat{H}, \hat{\Pi}])^2} &= -\expval{(\hat{H} \hat{\Pi} - \hat{\Pi} \hat{H})^2} \\
        &= \expval{\hat{H} \hat{\Pi} \hat{H}}.
    \end{align}
\end{subequations}
To derive a second formula similar to the case $\hat{\Pi} \ket{\psi} = \ket{\psi}$, we assume that $\ket{\psi}$ is an eigenvector of $\hat{P}$: $\hat{P} \ket{\psi} = \theta \ket{\psi}$, with $\theta \neq 1$ an $n$th root of unity. Then
\begin{equation}
    \frac{1}{n^2} \Delta\left(\sum_{l=0}^{n-1} \theta^l \hat{P}^l \hat{H} \hat{P}^{-l}\right) = \expval{\hat{H} \hat{\Pi} \hat{H}}.
\end{equation}
Note that for $\theta \neq -1$ (the only real non-trivial root), the operator $\sum_l \theta^l \hat{P}^l \hat{H} \hat{P}^{-l}$ is not necessarily Hermitian. Hence, only the case $\theta = -1$ (possible when $n$ is even) yields a physically meaningful variance. Using the eigenvalue condition:
\begin{equation}
    \expval{\sum_{l=0}^{n-1} \theta^l \hat{P}^l \hat{H} \hat{P}^{-l}} = \sum_{l=0}^{n-1} \theta^l \theta^{-l} \theta^{-l} \expval{\hat{H}} = \expval{\hat{H}} \sum_{l=0}^{n-1} \theta^{-l} = 0,
\end{equation}
and
\begin{subequations}
    \begin{align}
        \expval{\left(\sum_{l=0}^{n-1} \theta^l \hat{P}^l \hat{H} \hat{P}^{-l}\right)^2} &= \sum_{k,l=0}^{n-1} \theta^{k+l} \expval{ \hat{P}^k \hat{H} \hat{P}^{l-k} \hat{H} \hat{P}^{-l} } \\
        &= \sum_{k,l=0}^{n-1} \theta^0 \expval{ \hat{H} \hat{P}^{k-l} \hat{H} } \\
        &= n \sum_{l=0}^{n-1} \expval{\hat{H} \hat{P}^l \hat{H}} = n^2 \expval{\hat{H} \hat{\Pi} \hat{H}}.
    \end{align}
\end{subequations}

\section{Extension to mixed states}
In this section, we explain how the various results presented in the paper can be extended to the case of pure states. More explicitly, we first describe how the symmetry of a mixed state should be measured, then we explain how Eq.~(\ref{eq: PC HOM single photon}) and Eq.~(\ref{eq: proba fourier}) can be extended, showing how the photon-number distribution at the output of the interferometer is related to the symmetry of the mixed input states. Finally, we show that, given an initial mixed input state with the required symmetry property, our protocol provides an explicit measurement scheme for which we can evaluate the precision and assess optimality. We thus consider a mixed state $\hat \rho$ with the pure-state decomposition $\hat\rho=\sum_j p_j \ketbra{\psi_j}$ (with all $p_j\neq 0$). This decomposition can be thought of as a diagonalization of the density matrix (with $\braket{\psi_j}{\psi_k}=\delta_{j,k}$), but our results hold for any decomposition, not necessarily with orthogonal pure states.

First, the natural extension of the measure of symmetry $\bra{\psi}\hat S\ket{\psi}$ (for $n=2$) or $\bra{\psi}\hat P\ket{\psi}$ (for general $n$) is obtained by taking the mixed-state expectation values
\begin{align}
    \tr(\hat \rho\hat S) && \text{or} &&\tr(\hat \rho\hat P).
\end{align}
As usual, for the expectation value of a mixed state, we have
\begin{equation}
    \tr(\hat \rho\hat P)=\sum_j p_j\tr(\ketbra{\psi_j}\hat P)=\sum_j p_j \bra{\psi_j}\hat P\ket{\psi_j},
\end{equation}
so that $\tr(\hat \rho\hat P)$ can be interpreted as the average symmetry of the pure components of $\rho$ weighted by the probabilities $p_j$. An important case is when $\tr(\hat \rho \hat P)$ takes extremal values, {\it i.e.}, $\abs{\tr(\hat \rho\hat P)}=1$. In this case, let us write $\tr(\hat \rho\hat P)=e^{i\phi}$. It is then natural to expect that all pure-state components of $\hat\rho$ satisfy $\bra{\psi_j}\hat P\ket{\psi_j}=e^{i\phi}$. This can be formally proven by a clever use of the Cauchy-Schwarz inequality
\begin{equation}
    1=\abs{\sum_j p_j \bra{\psi_j}\hat P\ket{\psi_j}}^2\leq \sum_j p_j \sum_j p_j \abs{\bra{\psi_j}\hat P\ket{\psi_j}}^2=1,
\end{equation}
where we have used $\sum_j p_j=1$ and $\abs{\bra{\psi_j}\hat P\ket{\psi_j}}\leq 1$ since the eigenvalues of $\hat P$ have unit modulus. The condition for equality in the Cauchy-Schwarz inequality implies that the following vectors are proportional
\begin{equation}
    \begin{pmatrix}
        \sqrt{p_1}\\
        \vdots\\
        \sqrt{p_n}
    \end{pmatrix}\propto \begin{pmatrix}
        \sqrt{p_1}\bra{\psi_1}\hat P\ket{\psi_1}\\
        \vdots\\
        \sqrt{p_n}\bra{\psi_n}\hat P\ket{\psi_n}
    \end{pmatrix}.
\end{equation}
As all $p_j$ are assumed to be nonzero, this implies that all $\bra{\psi_j}\hat P\ket{\psi_j}$ are equal, necessarily to $e^{i\phi}$. An important consequence is that if $\tr(\hat \rho\hat S)= 1$ (resp. $-1$), then all pure states appearing in any decomposition of $\hat \rho$ are automatically symmetric (resp. antisymmetric).

If the state $\rho$ is injected into the DFT interferometer, the photon-number distribution in the output ports will simply be a weighted superposition of the distributions obtained for each pure component. As such, we have
\begin{equation}\label{eq: fourier proba mixed states}
    \frac{1}{n}\sum_{l=0}^{n-1}\Tr(\hat P^l\hat \rho)=\frac{1}{n}\sum_j p_j\sum_{l=0}^{n-1}\expval{\hat P^l}_{\ket{\psi_j}}=\sum_j p_j\mathbb P_{\ket{\psi_j}}\left[\sum_{k=0}^{n-1} k m_k\equiv0\pmod{n}\right]=\mathbb P_{\hat \rho}\left[\sum_{k=0}^{n-1} k m_k\equiv0\pmod{n}\right],
\end{equation}
where the notation $\mathbb P_{\hat \sigma}[\cdots]$ denotes the probability of a certain photon-number distribution when the state $\hat \sigma$ is injected into the DFT interferometer. This formula is directly analogous to Eq.~(\ref{eq: proba fourier}), linking the photon-number distribution to the symmetry properties of $\hat \rho$. In the case $n=2$, this equation reduces to
\begin{equation}
    \mathbb P_{\hat \rho}[n_1\equiv 0 \pmod{2}]=\frac{1}{2}\left(1+\tr(\hat \rho\hat S)\right),
\end{equation}
which is the natural extension of Eq.~(\ref{eq: PC HOM single photon}) to mixed states.

Finally, we consider the metrological problem of estimating $\kappa$ when the evolution $e^{i\hat H\kappa}$ is performed before the DFT interferometer with an initial mixed state. Since Eq.~(\ref{eq: fourier proba mixed states}) is linear in the quantum state, we can use the pure-state expansions of Eq.~(\ref{eq: expansion second order pure}) to develop the probability to second order in $\kappa$:
\begin{align}
    \mathbb{P}\left[\sum_{k=0}^{n-1} k m_k \equiv 0\;[n] \,\middle|\, \kappa\right] 
        &= \sum_j p_j \Bigg[\expval{\hat{\Pi}}_{\ket{\psi_j}} + i\kappa \left( \expval{\hat{\Pi} \hat{H}}_{\ket{\psi_j}} - \expval{\hat{H} \hat{\Pi}}_{\ket{\psi_j}} \right)\notag\\
        &\qquad- \tfrac{\kappa^2}{2} \left( \expval{\hat{H}^2 \hat{\Pi}}_{\ket{\psi_j}} + \expval{\hat{\Pi} \hat{H}^2}_{\ket{\psi_j}} - 2\expval{\hat{H} \hat{\Pi} \hat{H}}_{\ket{\psi_j}} \right) + o(\kappa^2)\Bigg].
\end{align}
Following the same reasoning as before, we require the constant term to be either $0$ or $1$, which means that $\tr(\hat \rho\hat \Pi)=1$ or $\tr(\hat \rho\hat \Pi)=0$. In both cases, using Eq.~(\ref{eq: second order for sym}) and Eq.~(\ref{eq: second order non sym}), we obtain
\begin{equation}
    \mathcal F=\sum_j p_j \Delta_{\ket{\psi_j}}(i[\hat H,\hat \Pi]).
\end{equation}
Considering Eq.~(\ref{eq: FI Fourier}), we can provide a more physically concrete formula if we make further assumptions on the symmetry of $\hat \rho$. If all pure components of $\hat\rho$ are symmetric, which is equivalent (as argued above) to $\tr(\hat \rho\hat P)=1$, we get
\begin{equation}
    \mathcal F=\frac{4}{n^2}\sum_j p_j \Delta_{\ket{\psi_j}}\left(n\hat H-\sum_{l=0}^{n-1} \hat P^l\hat H\hat P^{-l}\right).
\end{equation}
Conversely, if all pure components of $\hat\rho$ are antisymmetric, which is equivalent to $\tr(\hat \rho\hat P)=-1$, we have
\begin{equation}
    \mathcal F=\frac{4}{n^2}\sum_j p_j \Delta_{\ket{\psi_j}}\left(\sum_{l=0}^{n-1} (-1)^l\hat P^l\hat H\hat P^{-l}\right).
\end{equation}
Knowing that the quantum Fisher information (QFI) is convex, we obtain
\begin{align}\label{eq: ineq precision mixed state}
    \mathcal F\geq \mathcal Q\left(\hat H-\frac{1}{n}\sum_{l=0}^{n-1} \hat P^l\hat H\hat P^{-l}\right) &&\text{or} && \mathcal F\geq \mathcal Q\left(\frac{1}{n}\sum_{l=0}^{n-1}(-1)^l \hat P^l\hat H\hat P^{-l}\right),
\end{align}
depending on whether $\tr(\hat \rho\hat P)$ equals $1$ or $-1$, where $\mathcal Q(\hat H')$ denotes the QFI in the situation where the operator $\hat H'$ is used for the evolution instead of $\hat H$. As we now argue, we have equality in both equations above. Define $\hat H'$ as $\hat H-\frac{1}{n}\sum_{l=0}^{n-1} \hat P^l\hat H\hat P^{-l}$ or $\frac{1}{n}\sum_{l=0}^{n-1}(-1)^l \hat P^l\hat H\hat P^{-l}$, depending on whether we are considering the first or the second case. We can consider the alternative metrological situation where $\hat H'$ is used for the evolution instead of $\hat H$. Applying Eq.~(\ref{eq: ineq precision mixed state}) with $\hat H'$ instead of $\hat H$ bounds the FI as $\mathcal F(\hat H')\geq \mathcal Q(\hat H')$. Since the QFI gives the optimal precision obtainable, we necessarily have the reverse inequality $\mathcal F(\hat H')\leq \mathcal Q(\hat H')$. With a straightforward computation, we can verify that in both cases, $\mathcal F(\hat H')=\mathcal F(\hat H)$. Consequently, the precision of the protocol when the original generator $\hat H$ is used is
\begin{align}\label{eq: eq precision mixed state}
    \mathcal F= \mathcal Q\left(\hat H-\frac{1}{n}\sum_{l=0}^{n-1} \hat P^l\hat H\hat P^{-l}\right) &&\text{or} && \mathcal F= \mathcal Q\left(\frac{1}{n}\sum_{l=0}^{n-1}(-1)^l \hat P^l\hat H\hat P^{-l}\right),
\end{align}
depending on whether $\tr(\hat \rho\hat P)$ equals $1$ or $-1$. These formulas can be analyzed in the same spirit as those valid for pure states. In particular, if $\hat H$ is antisymmetric under cyclic permutation ($\hat P\hat H\hat P^\dagger=-\hat H$), the measurement device is optimal with $\mathcal F=\mathcal Q$.

\section{Analysis of the photon loss}
\color{red}
In this appendix, we present an approach to understanding the effect of photon losses. Throughout this section, we assume that photon loss occurs independently in each mode, with the same transmissivity $\eta$ for all modes. This model encompasses photon absorption, scattering, and imperfect detection efficiency. In practice, we represent losses using a lossy BS model, which transforms the creation operators as
\begin{equation}
    \hat a_j^\dagger(\lambda) \to \sqrt{\eta} \hat a_j^\dagger(\lambda) + \sqrt{1-\eta} \hat b_j^\dagger(\lambda)
\end{equation}
where $\hat b_j^\dagger(\lambda)$ denotes the creation operator of an environmental mode initially in the vacuum state.

\paragraph{$\blacktriangleright$ Position of the losses} 
Losses can occur at any stage of the experiment—before or after the evolution $\hat V(\kappa)$ and the interferometer $\hat U$. If they occur immediately before the measurement process, they simply modify the photon-number statistics. However, losses that occur earlier can alter the quantum state itself, potentially affecting its subsequent evolution and interference. Nevertheless, as we now argue, losses can be commuted with linear optical operations, meaning that the same output statistics are obtained whether loss occurs before or after a linear optical transformation. Consider the mode transformation $\hat M$ defined by
\begin{equation}
    \hat a_j^\dagger \mapsto \sum_{k=0}^{n-1} M_{k,j} \hat a_k^\dagger.
\end{equation}
Depending on the order of composition, the operator $\hat a_j^\dagger(\lambda)$ transforms as
\begin{subequations}
    \begin{align}
        \hat a_j^\dagger(\lambda) &\mapsto \sqrt{\eta} \sum_{k=0}^{n-1} M_{k,j} \hat a_k^\dagger(\lambda) + \sqrt{1-\eta} \hat b_j^\dagger(\lambda) && \text{(loss after } \hat M\text{),}\\
        \hat a_j^\dagger(\lambda) &\mapsto \sqrt{\eta} \sum_{k=0}^{n-1} M_{k,j} \hat a_k^\dagger(\lambda) + \sqrt{1-\eta} \sum_{k=0}^{n-1} M_{k,j} \hat b_k^\dagger(\lambda) && \text{(loss before } \hat M\text{).}
    \end{align}
\end{subequations}
In these two expressions, the only difference lies in the environmental modes: $\hat b_j^\dagger(\lambda)$ versus $\hat c_j^\dagger(\lambda)=\sum_{k=0}^{n-1} M_{k,j}\hat b_k^\dagger(\lambda)$. However, since these modes are unobserved, the distinction is irrelevant: the particular mode in which photons are lost does not affect the measurable output statistics. Assuming that the environmental modes are initially in the vacuum state, we can compute the output probabilities for a general input state $\ket{\psi}$ by expanding the transformed creation operators. Because of the equivalence between the environmental modes in the two scenarios, the output statistics are identical. Hence, losses commute with linear optical operations without affecting the measurement outcomes. 

Since the interferometer $\hat U$ is, by definition, a linear optical transformation, we can equivalently assume that loss occurs either before or after $\hat U$. Similarly, in many cases, such as phase estimation or delay measurement, the evolution operator $\hat V(\kappa)$ is also linear. In such situations, all losses can be commuted to the beginning of the experiment, effectively acting as imperfect state preparation.
Therefore, we can consider losses as occurring at any point in the experimental setup without changing the output statistics.

\medskip
\paragraph{$\blacktriangleright$ Linking ideal and lossy probabilities}
Having established that losses can be considered at the detection stage, we now relate the ideal and lossy probability distributions. As the observable counts the number of photons detected in each spatial mode, we define
\begin{align}
    P_{j_0,\dots,j_{n-1}} &= \mathbb{P}[m_0=j_0,\dots,m_{n-1}=j_{n-1}], &
    Q_{k_0,\dots,k_{n-1}} &= \mathbb{P}_\eta[m_0=k_0,\dots,m_{n-1}=k_{n-1}],
\end{align}
where $P_{j_0,\dots,j_{n-1}}$ denotes the ideal probability of detecting $j_l$ photons in mode $l$, and $Q_{k_0,\dots,k_{n-1}}$ the corresponding probability in the presence of losses. For a single mode ($n=1$), a simple combinatorial argument yields
\begin{equation}
    Q_k = \sum_{j=k}^\infty \binom{j}{k} \eta^k (1-\eta)^{j-k} P_j.
\end{equation}
This can be interpreted as follows: detecting $k$ photons after loss could originate from any initial number $j\geq k$ of photons, of which $k$ were transmitted and $j-k$ lost. The binomial coefficient counts the number of possible transmitted subsets, while the factors $\eta^k$ and $(1-\eta)^{j-k}$ represent the transmission and loss probabilities, respectively. Extending this reasoning to $n$ modes, we obtain
\begin{equation}
    Q_{k_0,\dots,k_{n-1}} = \sum_{j_0=k_0}^\infty \cdots \!\!\!\sum_{j_{n-1}=k_{n-1}}^\infty \prod_{l=0}^{n-1} \binom{j_l}{k_l} \eta^{k_l}(1-\eta)^{j_l-k_l} P_{j_0,\dots,j_{n-1}}.
\end{equation}

In practice, one typically measures $Q_{k_0,\dots,k_{n-1}}$ and wishes to recover the ideal probabilities $P_{j_0,\dots,j_{n-1}}$. For $n=1$, this inverse relation can be derived conveniently using generating functions. Define
\begin{align}
    F(x) = \sum_{k=0}^\infty P_k x^k, \qquad G(x) = \sum_{k=0}^\infty Q_k x^k.
\end{align}
From the relation above, we obtain
\begin{subequations}
    \begin{align}
        G(x) &= \sum_{k=0}^\infty Q_k x^k\\
        &= \sum_{k=0}^\infty \sum_{j=k}^\infty \binom{j}{k} (1-\eta)^k \eta^{j-k} P_i x^k\\
        &= \sum_{j=0}^\infty \sum_{k=0}^j \binom{j}{k} (1-\eta)^k \eta^{j-k} P_j x^k\\
        &= \sum_{j=0}^\infty \big[\eta + x(1-\eta)\big]^j P_i\\
        &= F\big(\eta + x(1-\eta)\big).
    \end{align}
\end{subequations}
Thus, the inverse map $x \mapsto \frac{x-\eta}{1-\eta}$ gives
\begin{subequations}
    \begin{align}
        F(x) &= G\left(\frac{x-\eta}{1-\eta}\right)\\
        &= \sum_{j=0}^\infty \left(\frac{x-\eta}{1-\eta}\right)^j Q_j\\
        &= \sum_{j=0}^\infty \frac{1}{(1-\eta)^j} \sum_{k=0}^j \binom{j}{k} (-\eta)^{j-k} x^k Q_j\\
        &= \sum_{k=0}^\infty \sum_{j=k}^\infty \binom{j}{k} \frac{(-\eta)^{j-k}}{(1-\eta)^j} Q_j x^k.
    \end{align}
\end{subequations}
By identifying coefficients, we recover
\begin{equation}
    P_k = \sum_{j=k}^\infty \binom{j}{k} \frac{(-\eta)^{j-k}}{(1-\eta)^j} Q_j.
\end{equation}
Extending this to $n$ modes yields
\begin{equation}
    P_{j_0,\dots,j_{n-1}} = \sum_{k_0=j_0}^\infty \cdots \!\!\!\sum_{k_{n-1}=j_{n-1}}^\infty \prod_{l=0}^{n-1} \binom{k_l}{j_l} \frac{(-\eta)^{k_l-j_l}}{(1-\eta)^{k_l}} Q_{k_0,\dots,k_{n-1}}.
\end{equation}
This relation allows one to reconstruct the ideal probabilities from the measured lossy ones, thereby enabling the computation of the ideal probability $\mathbb P[\sum_{k=0}^{n-1} k m_k \equiv 0;[n]]$ required for parameter estimation. Furthermore, using the formal expression of the Fisher information, one can quantify the efficiency of estimation in the presence of losses.

\medskip
\paragraph{$\blacktriangleright$ Symmetry of lossy single photon states}
As argued in the main text, the symmetry of the input state plays a crucial role in the applicability of our methods and the success of the estimation procedure. Since we have shown that losses can be commuted to the state-preparation stage (when $\hat V(\kappa)$ is an optical linear operation), losses are effectively equivalent to imperfect state preparation. This motivates an analysis of the relationship between loss and state symmetry. For single-photon states, an interesting structure emerges. Recall that such a state can be written as
\begin{equation}
    \ket{\psi} = \int \dd\lambda_0 \cdots \dd\lambda_{n-1} F(\lambda_0,\dots,\lambda_{n-1}) \hat a_0^\dagger(\lambda_0)\cdots \hat a_{n-1}^\dagger(\lambda_{n-1}) \vac.
\end{equation}
Under the loss model, this state transforms into
\begin{equation}
    \ket{\psi_\eta} = \int \dd\lambda_0 \cdots \dd\lambda_{n-1} F(\lambda_0,\dots,\lambda_{n-1}) \prod_{j=0}^{n-1} \left(\sqrt{\eta} \hat a_j^\dagger(\lambda_j) + \sqrt{1-\eta} \hat b_j^\dagger(\lambda_j)\right)\vac.
\end{equation}

To manipulate this expression, let us introduce a convenient notation.
For a binary string $s \in \{a,b\}^n$, define $\ket{\psi_s}$ as the state obtained from $\ket{\psi}$ by replacing $\hat a_j^\dagger(\lambda_j)$ with $\hat b_j^\dagger(\lambda_j)$ whenever the $j$th letter of $s$ is $b$, and leaving it unchanged otherwise. For example, for $n=3$ and $s=aba$,
\begin{equation}
    \ket{\psi_{aba}} = \int \dd\lambda_0 \dd\lambda_1 \dd\lambda_2 F(\lambda_0,\lambda_1,\lambda_2) \hat a_0^\dagger(\lambda_0) \hat b_1^\dagger(\lambda_1) \hat a_2^\dagger(\lambda_2)\vac.
\end{equation}
We also have $\ket{\psi} = \ket{\psi_{a\cdots a}}$. Using this notation,
\begin{equation}
    \ket{\psi_\eta} = \sum_{s\in\{a,b\}^n} \eta^{n_a(s)/2} (1-\eta)^{n_b(s)/2} \ket{\psi_s},
\end{equation}
where $n_a(s)$ and $n_b(s)$ count the occurrences of $a$ and $b$ in $s$, respectively. After tracing out the environmental ($b$) modes, we obtain
\begin{equation}
    \hat \rho_\eta = \sum_{s,t\in\{a,b\}^n} \eta^{\frac{n_a(s)+n_a(t)}{2}} (1-\eta)^{\frac{n_b(s)+n_b(t)}{2}} \tr_b(\ketbra{\psi_s}{\psi_t}).
\end{equation}
Since $\ket{\psi_s}$ and $\ket{\psi_t}$ differ in at least one mode whenever $s\neq t$, we have $\tr_b(\ketbra{\psi_s}{\psi_t})=0$ for $s\neq t$. Thus, only the diagonal terms contribute
\begin{equation}
    \hat \rho_\eta = \sum_{s\in\{a,b\}^n} \eta^{n_a(s)} (1-\eta)^{n_b(s)} \tr_b(\ketbra{\psi_s}).
\end{equation}
The average symmetry of the resulting state is therefore
\begin{equation}
    \operatorname{Tr}(\hat \rho_\eta \hat P) = \sum_{s\in\{a,b\}^n} \eta^{n_a(s)} (1-\eta)^{n_b(s)} \tr(\tr_b(\ketbra{\psi_s}{\psi_s}) \hat P).
\end{equation}
Here, $\tr_b(\ketbra{\psi_s}{\psi_s})$ contains a photon in spatial mode $j$ only if $s_j=a$. Since $\hat P$ cyclically permutes the spatial modes, the trace above is nonzero only if $s$ is invariant under cyclic permutation. The only such strings are $s=a\cdots a$ and $s=b\cdots b$. Hence,
\begin{equation}
    \tr(\hat \rho_\eta \hat P) = \eta^n \bra{\psi}\hat P\ket{\psi} + (1-\eta)^n.
\end{equation}
For a mixed single-photon state $\hat \rho$, this generalizes to
\begin{equation}
    \tr(\hat \rho_\eta \hat P) = \eta^n \operatorname{Tr}(\hat \rho \hat P) + (1-\eta)^n.
\end{equation}
\color{black}

\fi

\ifnum 2=\theShow
    \bibliographystyle{unsrturl} 
    \bibliography{refs}
\fi

\end{document}